\documentclass{emulateapj}

\usepackage{xspace}
\usepackage{graphics,graphicx}
\usepackage{natbib}
\usepackage{multirow}
\usepackage{longtable}
\usepackage{rotating}
\usepackage{amsmath}
\usepackage[percent]{overpic}
\usepackage{xcolor}

\newcommand{\asn}{$^{\prime\prime}$\xspace}

\newcommand{\nH}{$N_{\rm H}$\xspace}
\newcommand{\PL}{$\Gamma$\xspace}
\newcommand{\Msun}{$M_{\odot}$\xspace}

\newcommand{\Chandra}{{\it Chandra}\xspace}
\newcommand{\lum}{erg s$^{-1}$\xspace}
\newcommand{\flux}{erg s$^{-1}$ cm$^{-2}$\xspace}

\newcommand{\lognlogs}{log$N$-log$S$\xspace}

\shorttitle{Multiple Epochs of the NGC~300 \lognlogs Distribution}
\shortauthors{Binder et al.}

\begin{document}
 
\title{The Effect of Variability on X-Ray Binary Luminosity Functions: Multiple Epoch Observations of NGC 300 with {\it Chandra}}
\author{B. Binder\altaffilmark{1,2}, 
J. Gross\altaffilmark{1}, 
B. F. Williams\altaffilmark{1},
M. Eracleous\altaffilmark{3},
T. J. Gaetz\altaffilmark{4},
P. P. Plucinsky\altaffilmark{4},
E. D. Skillman\altaffilmark{5}
}
\altaffiltext{1}{University of Washington, Department of Astronomy, Box 351580, Seattle, WA 98195}
\altaffiltext{2}{Department of Physics \& Astronomy, California State Polytechnic University, 3801 West Temple Ave, Pomona, CA 91768}
\altaffiltext{3}{Department of Astronomy \& Astrophysics, The Pennsylvania State University, 525 Davey Lab, University Park, PA 16802}
\altaffiltext{4}{Harvard-Smithsonian Center for Astrophysics, 60 Garden Street Cambridge, MA 02138}
\altaffiltext{5}{Minnesota Institute for Astrophysics, University of Minnesota, 116 Church St. SE, Minneapolis, MN 55455}

\begin{abstract}
We have obtained three epochs of \Chandra ACIS-I observations (totaling $\sim$184 ks) of the nearby spiral galaxy NGC~300 to study the \lognlogs distributions of its X-ray point source population down to $\sim$2$\times$10$^{-15}$ \flux in the 0.35-8 keV band (equivalent to $\sim$10$^{36}$ \lum). The individual epoch \lognlogs distributions are best described as the sum of a background AGN component, a simple power law, and a broken power law, with the shape of the \lognlogs distributions sometimes varying between observations. The simple power law and AGN components produce a good fit for ``persistent'' sources (i.e., with fluxes that remain constant within a factor of $\sim$2). The differential power law index of $\sim$1.2 and high fluxes suggest that the persistent sources intrinsic to NGC~300 are dominated by Roche lobe overflowing low mass X-ray binaries. The variable X-ray sources are described by a broken power law, with a faint-end power law index of $\sim$1.7, a bright-end index of $\sim$2.8--4.9, and a break flux of $\sim$8$\times10^{-15}$ \flux ($\sim$4$\times10^{36}$ \lum), suggesting they are mostly outbursting, wind-fed high mass X-ray binaries, although the \lognlogs distribution of variable sources likely also contains low-mass X-ray binaries. We generate model \lognlogs distributions for synthetic X-ray binaries and constrain the distribution of maximum X-ray fluxes attained during outburst. Our observations suggest that the majority of outbursting X-ray binaries occur at sub-Eddington luminosities, where mass transfer likely occurs through direct wind accretion at $\sim$1--3\% of the Eddington rate. 
\end{abstract}

\section{Introduction}
X-ray binaries (XRBs) are a nearly ubiquitous constituent of galaxies \citep{Shty+07}, and X-ray luminosity functions (XLFs) of XRBs have become a standard tool for investigating their characteristics across a range of environments. The XRB population of star-forming galaxies is dominated by high-mass systems (HMXBs), whose XLFs follow a ``universal'' power law with a cumulative slope of $\sim$0.6 \citep{Kilgard+02,Grimm+03,Mineo+12}. The shape of the XLF is remarkably uniform, both across galaxies and multiple epochs of individual galaxies \citep[e.g., as observed in the Antennae;][]{Zezas+07}, down to limiting luminosities of $\sim$10$^{37}$ \lum. However, individual HMXBs have been observed to exhibit high levels of variability, both in X-ray spectral shape and luminosity, over timescales ranging from minutes to years \citep{Reig08}. No progress has yet been made on reconciling the X-ray variability of individual sources with the stability of the population-wide XLF, likely due to the lack of observations at fainter luminosities ($<$10$^{37}$ \lum) where most XRBs are expected to be found.

We use NGC~300 \citep[at a distance of 2.0 Mpc, ][]{Dalcanton+09} as a laboratory for studying the effects of low-luminosity variability on the shape of the XLF. NGC~300 has had enough recent star formation to produce a large population of X-ray  sources \citep[nearly one hundred discrete X-ray sources have been detected down to a 0.35-8 keV luminosity of 10$^{36}$ \lum;][]{Binder+12} while producing only minimal diffuse X-ray emission. The galaxy is close enough so that faint HMXBs can be detected in reasonable exposure times but far enough away that the entire star-forming disk can be imaged in a single \Chandra exposure. Due to its isolation, the star forming disk of NGC~300 is relatively undisturbed, with no evidence of a merger event for the last $\sim$6 Gyr \citep{Bland-Hawthorn+05,Gogarten+10}.

We have obtained three epochs of \Chandra imaging of NGC~300, totaling $\sim$184 ks, to study discrete X-ray point source variability and its effects on the shape of the XLF. In Section~\ref{obs}, we present our observations and data reduction procedures. In Section~\ref{xlfs}, we construct and model the observed \lognlogs distributions of the three epochs (which can be converted to an XLF when all sources are assumed to lie at the same distance from the observer). In Section~\ref{discussion}, we demonstrate how the observed \lognlogs distributions can be reproduced from a population of individually variable sources, and discuss implications for XRB evolution. We conclude with a summary of our findings in Section~\ref{conclusions}.

\section{Observations and Data Reduction}\label{obs}

\begin{figure*}[ht]
\centering
\begin{tabular}{cc}
	\begin{overpic}[width=0.4\linewidth,clip=true,trim=0cm 0cm 0cm 0cm]{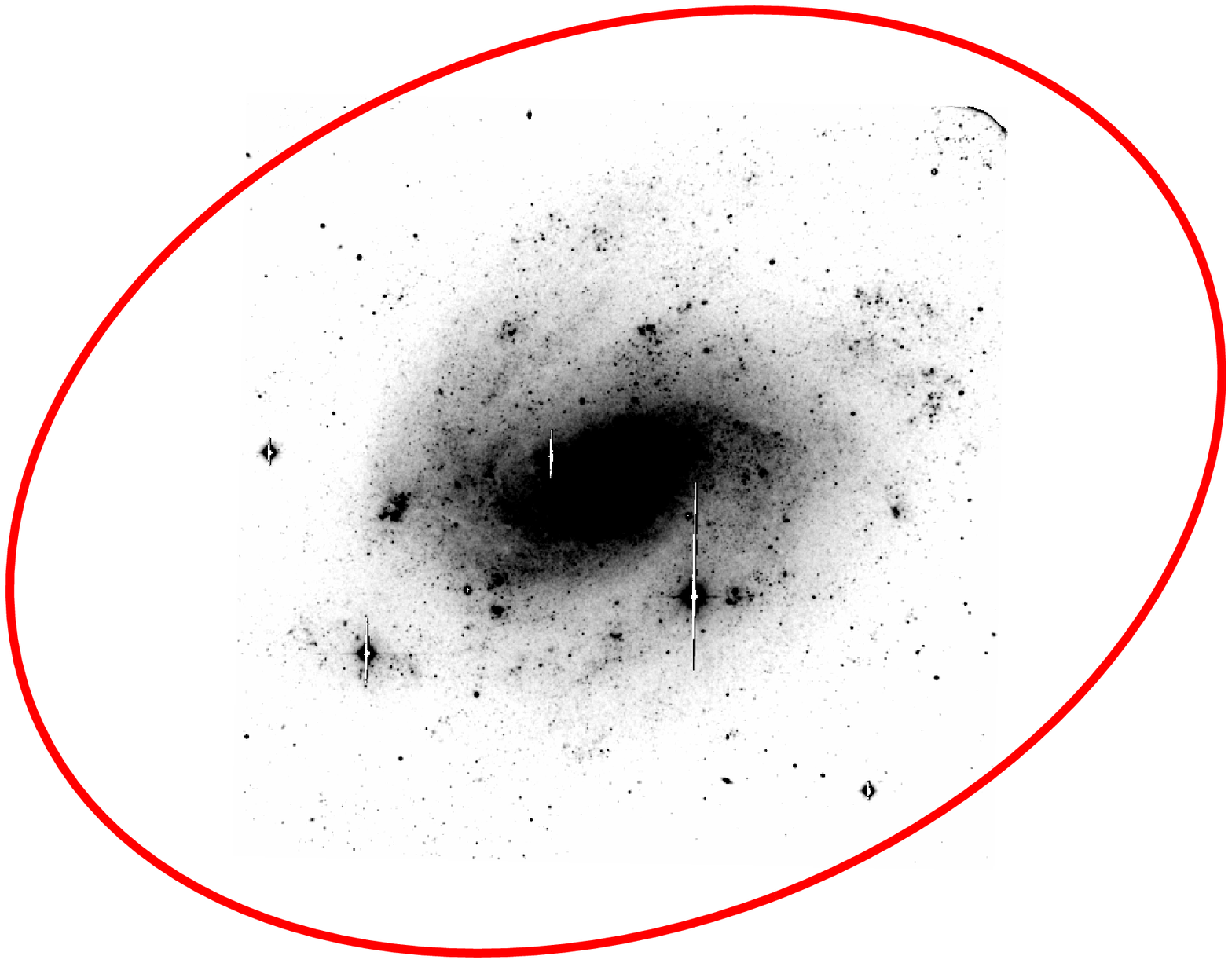}
 		\put (-2,93){\color{black}\normalsize ground-based, R-band}
	\end{overpic} &
	\begin{overpic}[width=0.4\linewidth]{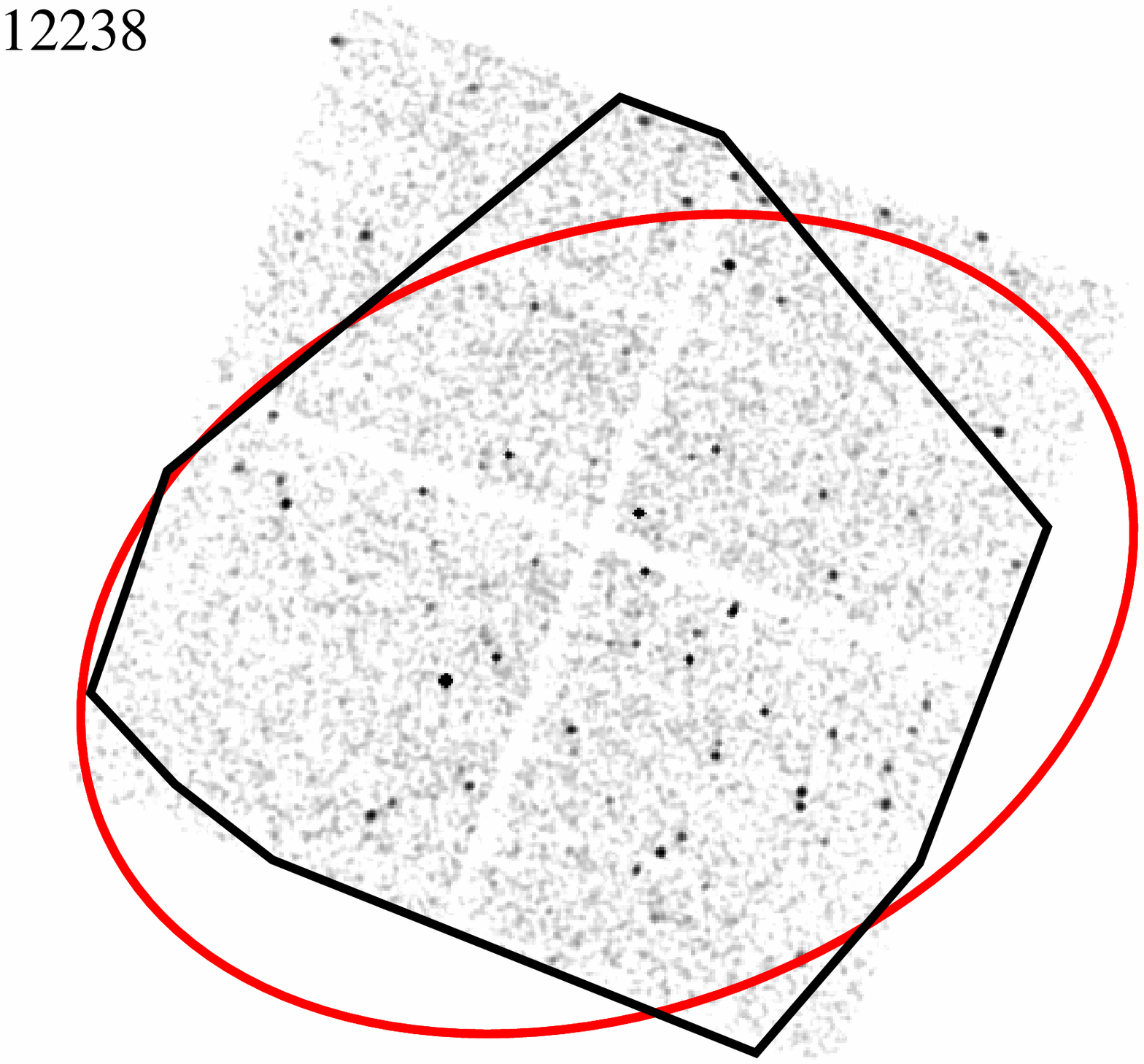}
	
	\end{overpic} \\

	\begin{overpic}[width=0.4\linewidth]{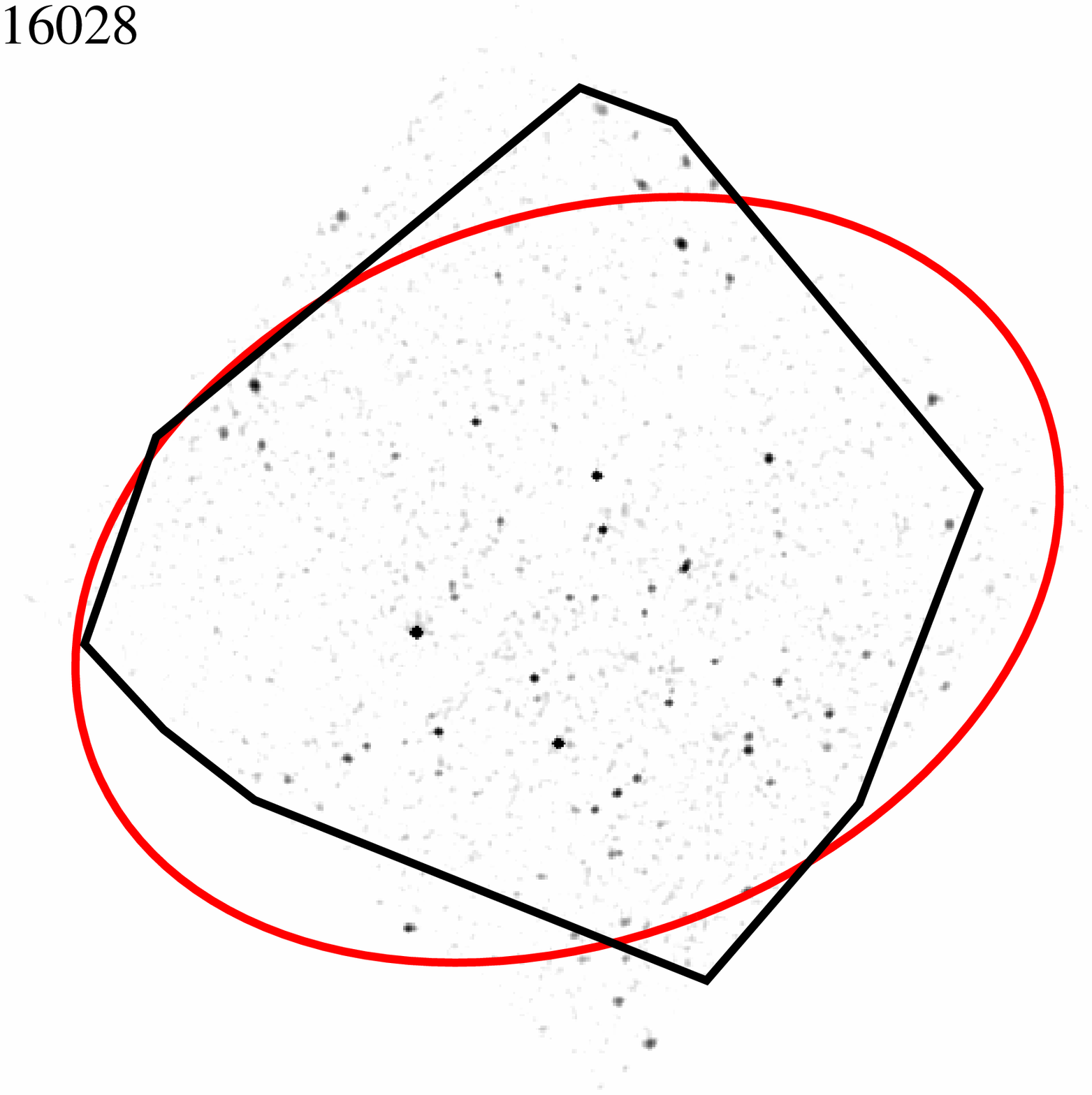}
	\end{overpic} &

	\begin{overpic}[width=0.4\linewidth]{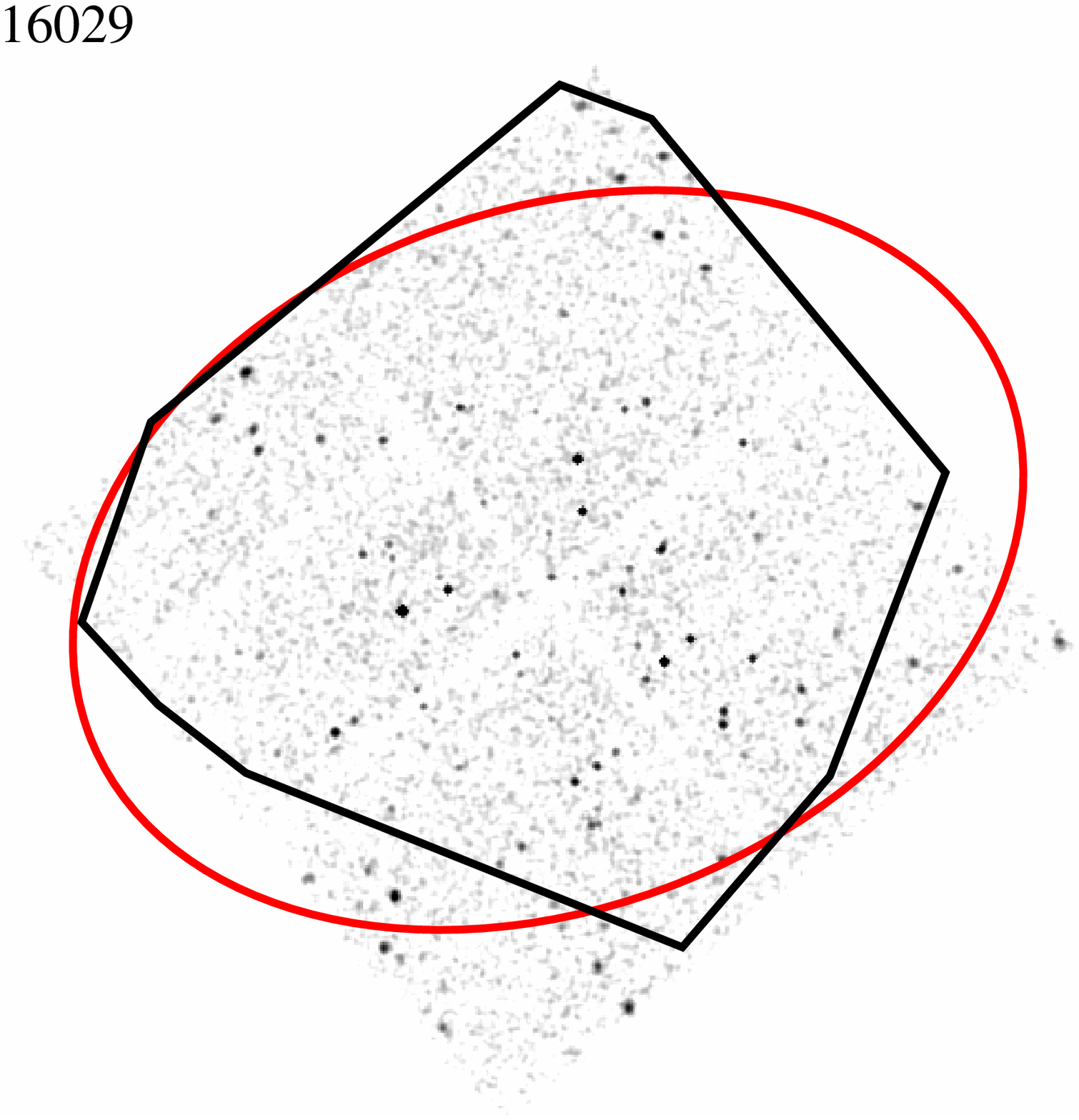}
	\end{overpic} \\
\end{tabular}
\caption{{\it Top left}: a ground-based, $R$-band image of NGC~300 \citep[obtained from the NASA/IPAC Extragalactic Database;][]{Larsen+99}. {\it Top right and bottom row}: \Chandra 0.35-8 keV images, with the corresponding ObsID indicated in the upper-left corner. The ``common area'' of our survey is outlined in black. In all images, the red ellipse shows the $R$-band 25 mag arcsec$^{-2}$ isophote. An animated version of this figure is provided by the journal.}\label{figure_merged}
\end{figure*}

We have observed NGC~300 three times with the \Chandra ACIS-I instrument; the observation identification numbers (hereafter referred to as the ``ObsIDs''), dates of the observations, and useable exposure times are summarized in Table~\ref{table_xray_obs}. Data reduction was carried out with CIAO v4.8 and CALDB v4.6.1.1 using standard reduction procedures\footnote{See \url{http://asc.harvard.edu/ciao/threads/index.html}}. A full analysis of ObsID 12238 was presented in \cite{Binder+12}; however, for consistency all three data sets were reprocessed using the CIAO task \texttt{chandra\_repro}. Exposure maps were constructed using the CIAO tool \texttt{flux\_obs}, which produces exposure-corrected images using user-specified instrument maps. For our instrument maps, we assumed spectral weights appropriate for both XRBs and AGN: a power-law spectrum (with \PL=1.7) absorbed by the average foreground column density \citep[\nH = 4.09$\times10^{20}$ cm$^{-2}$][]{Kalberla+05}. Background light curves were extracted and inspected for flares using the \texttt{lc\_clean} routine. No strong background flares were present in any of the exposures; the background light curves were clipped at 5$\sigma$ to create good time intervals (GTIs). All event data were filtered on the resulting GTIs. The exposures were corrected for (small) relative astrometric offsets using the CIAO tools \texttt{wcs\_match} and \texttt{wcs\_update}, and a single ``merged'' events file was created using \texttt{reproject\_obs}.

\begin{table}[ht]
\centering
\caption{Observation Log}
\begin{tabular}{ccc}
\hline \hline
\multirow{2}{*}{Obs. ID} & \multirow{2}{*}{Date}	& Exposure 	\\
			&						& Time (ks)	\\
 (1)			& (2)						& (3)      		\\
\hline	
12238		& 2010 Sept. 24			& 63.0		\\
16028		& 2014 May 16-17			& 63.9		\\
16029		& 2014 Nov. 17-18 			& 61.3		\\
\hline \hline
\end{tabular}
\label{table_xray_obs}
\end{table}

Figure~\ref{figure_merged} shows a ground-based $R$-band image of NGC~300 \citep[obtained from the NASA/IPAC Extragalactic Database; ][]{Larsen+99} and our three 0.35-8 keV \Chandra observations. There is no evidence for soft, diffuse X-ray emission (e.g., from hot gas, although ACIS-I is less sensitive to emission below 1.0 keV than ACIS-S), and numerous X-ray point sources are visible. The black outline shows the ``common area'' of our three observations which is used in the remainder of our analysis, and the red ellipse shows the $R$-band 25 mag arcsec$^{-2}$ isophote for reference. An animation of our three observations, provided by the journal, makes variable X-ray sources easily visible by eye.

		\subsection{Point Source Detection}\label{section_detect}
The CIAO task \texttt{wavdetect} \citep{Freeman+02} is a wavelet algorithm for \Chandra observations that is capable of separating even moderately crowded sources. We use \texttt{wavdetect} to perform point source detection on all three observations individually and on the merged image. On each image, we use \texttt{scales} of 1\asn, 2\asn, 4\asn, 8\asn and 16\asn in three different energy bands (0.35-8 keV, 0.35-2 keV, and 2-8 keV) with three different binning schemes (binned to 1, 4, and 9 pixels). The \texttt{sigthresh} parameter, the threshold for identifying a pixel as belonging to a source, was set to 6$\times10^{-8}$ (approximately one divided by the number of pixels in the merged image). The \texttt{bkgsigthresh} parameter, the statistical criterion for rejecting the null hypothesis that the pixel in question is due solely to the background, was set to $10^{-3}$. 

The resulting source lists were merged, keeping only unique source positions, and sources were visually examined for possible false detections. Spurious point sources, such as those observed at large off-axis angles with distorted point-spread functions (PSFs) that were split into two or more sources and sources with zero size, were removed from our source list. Only sources with a \texttt{wavdetect} significance $\sigma>3.5$ in at least one of the four images (the three individual exposures or the merged image) were included in our final catalog. Nine sources were detected in the merged image at $\sigma>3.5$, but not in any individual exposure. These sources were not included in our analysis, as their variability properties are unconstrained.

The final source list contains 115 X-ray point sources. Since the \texttt{wavdetect} algorithm can give slightly offset centroids in different images for the same source, the final position of each source that was detected in multiple images was derived by averaging the positions of the individual detections weighted by detection significance, as was done in \cite{Liu11}. For consistency with \cite{Liu11}, we use the empirical equation from \citet[][their section 5]{Kim+04x} to estimate the positional uncertainty as a function of off-axis angle on the ACIS-I detector and the number of net counts for each source. 

For each individual exposure, the photon counts were computed by fitting each source image to a 2D Gaussian. To define the ``source region,'' we first found the elliptical region that contained 95\% of the source counts for a Gaussian distribution. The semi-major and semi-minor axes were then increased by $\sim$20\%. We define a background annulus with an inner radius set to the semi-major axis of the source region. The outer radius of the annulus was determined such that the background region contained at least 50 counts. Radial surface brightness profiles were extracted and visually examined for each source, and source and background regions were adjusted (e.g., made more circular or elliptical, or by masking nearby sources) so that they did not contain other nearby point sources, residual source counts, etc.

		\subsection{Sensitivity Maps}
To construct an XLF, a sensitivity map providing the number of counts above which a source would be detectable at each point in our survey area is required. Sensitivity maps were made using the CIAO task \texttt{lim\_sens} for all three individual exposures. Count rates were then converted to energy fluxes assuming a power law with \PL = 1.7 obscured only by the Galactic absorbing column along the line of sight to NGC~300 \citep{Kalberla+05}. Since the exposure times, pointings, and instruments are nearly identical in all three observations, the resulting sensitivity limits are nearly identical in all three exposures. Furthermore, because each X-ray source is imaged at a similar location on the ACIS-I detector in each observation, there is no additional systematic uncertainty due to sensitivity variations across the detector. 

In the 0.35-8 keV band, we find that 90\% of the sensitivity map area has a flux value above 2$\times10^{-15}$ \flux (corresponding to a luminosity of 10$^{36}$ \lum at the distance of NGC~300) in our shallowest exposure. The sensitivity maps reach flux values of 10$^{-15}$ \flux and $4\times10^{-15}$ \flux over 75\% and 99\% of the map area, respectively (corresponding to respective luminosities of 6$\times10^{35}$ \lum and 2$\times10^{36}$ \lum). These luminosities are only $\sim$4\% fainter for our deepest exposure. This luminosity limit is about an order of magnitude fainter than what was reached in the variability study of the Antennae \citep{Zezas+07}. To avoid issues related to exposure time variations, we restrict all subsequent analysis to fluxes above $2\times10^{-15}$ \flux (e.g., the 90\% completeness limit). In all three observations, a difference of one net count corresponds to a change in unabsorbed 0.35-8 keV flux of $\sim2.5\times10^{-16}$ \flux (corresponding to $\sim1.2\times10^{35}$ \lum).

\section{The log$N$-log$S$ Distributions}\label{xlfs}
We calculate the \lognlogs distributions for each of our three observations, using the 85 X-ray sources that were detected in the common area of each observation. The cumulative number of sources per deg$^2$, $N$, above a given flux limit $S$ (in units of \flux) can be computed as

\begin{equation}
N(>S) = \sum_{i} \frac{1}{A(S_i)} \text{ deg}^{-2},
\end{equation}

\noindent where $A$ is the geometric area of the survey over which the $i$th source with a flux $S_i$ could be detected. Multiplying by the common area (0.07654 deg$^2$) yields the expected number of sources within the field of view. The 0.35-8 keV sensitivity maps allow us to directly evaluate the area function for each source in our survey. The standard deviation $\sigma$ in the number of sources in each flux bin ($n$) is estimated using the \cite{Gehrels86} approximations for upper limits,

\begin{equation}
\sigma_{\rm up} = 1 + \sqrt{n + 0.75},
\end{equation}

\noindent and lower limits,

\begin{equation}
\sigma_{\rm lo} = \sqrt{n - 0.25}.
\end{equation}

Once the cumulative \lognlogs distribution is computed, the differential \lognlogs distribution may be calculated as:

\begin{equation}
\frac{dN}{dS} = \frac{N(>[S + \Delta S]) - N(>S)}{\Delta S}.
\end{equation}

\noindent We use a bin size $\Delta S$ of 2.5$\times10^{-16}$ \flux (corresponding to a difference of $\sim$1 net count in our observations) to calculate the differential \lognlogs distributions. If all X-ray sources were at the same distance from the observer, the \lognlogs distribution could be directly converted into an XLF. However, the observed point sources in our survey are a mix of both sources intrinsic to NGC~300 and background AGN, and are therefore not all at a common distance.

\begin{figure*}
\centering
\begin{tabular}{cc}
\includegraphics[width=0.4\linewidth]{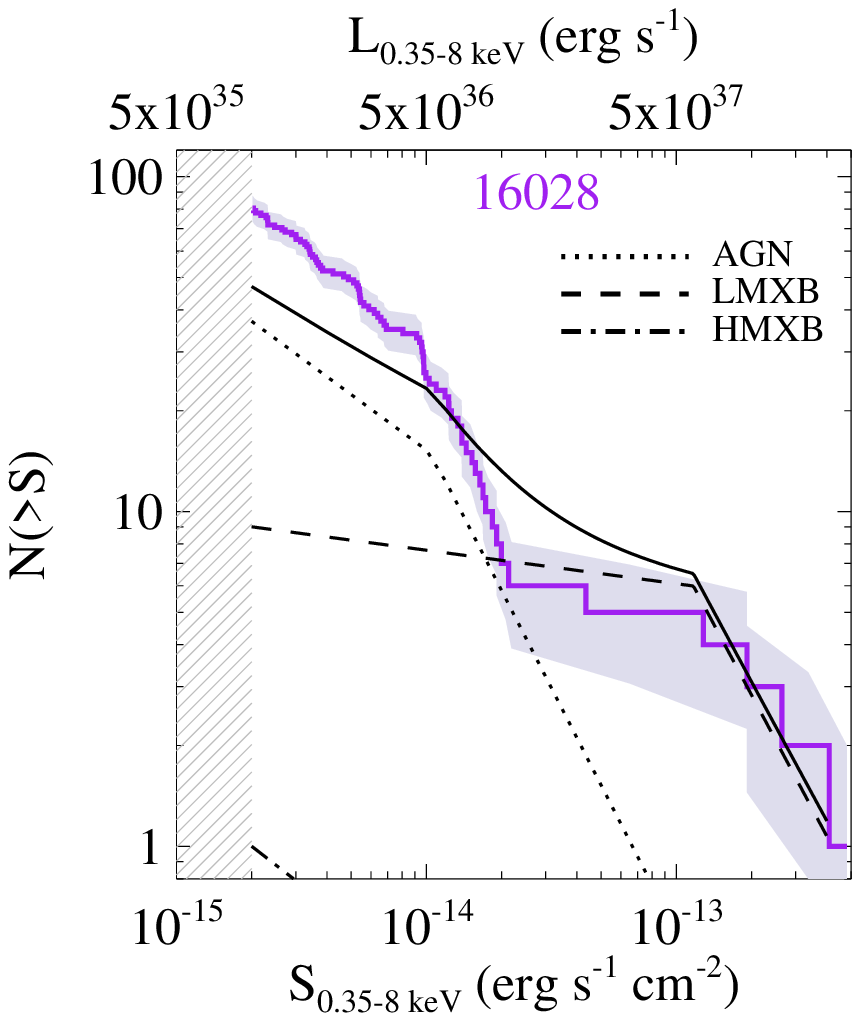} &
\includegraphics[width=0.4\linewidth]{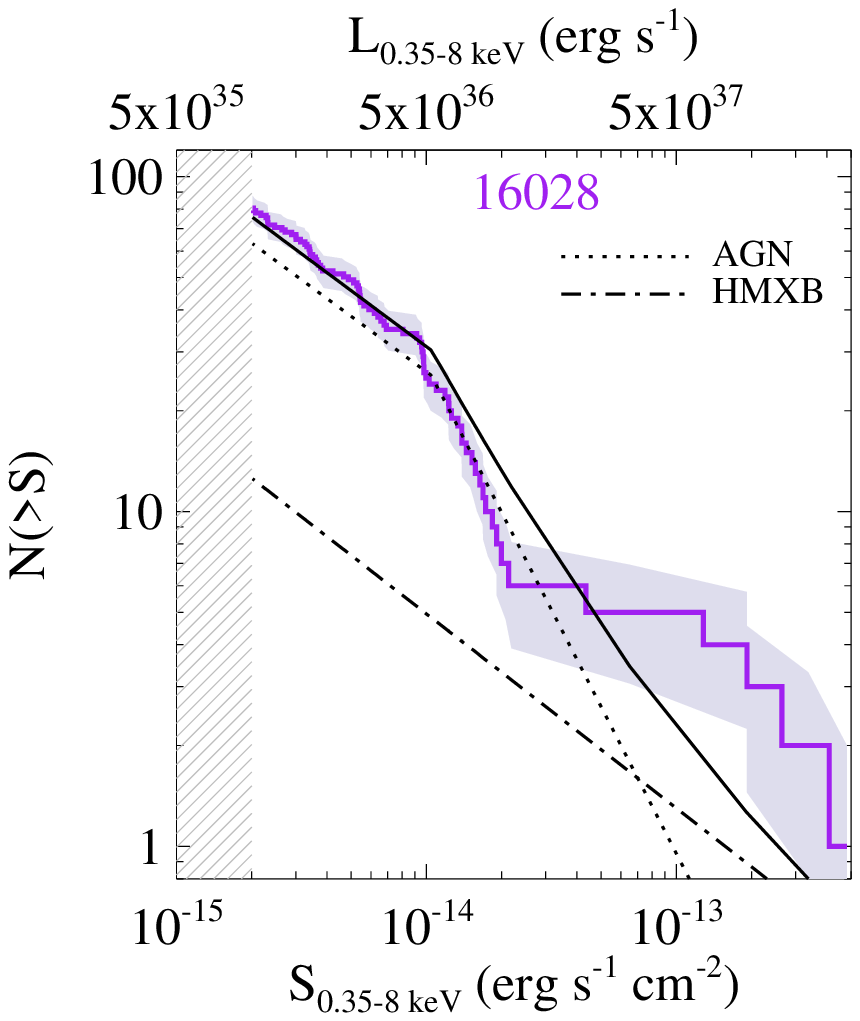} \\
\end{tabular}
\caption{The cumulative \lognlogs distribution for ObsID 16028. The shaded region shows the uncertainty in the number of observed sources. The gray-lined region indicates fluxes below our 90\% completeness limit; only sources above this limit are shown. The left panel shows the predicted AGN, HMXB, and LMXB contributions based on \lognlogs distributions in the literature. The right panel shows the resulting fit when the normalizations are left as free parameters (e.g., the LMXB normalization is consistent with zero). See Section~\ref{section_modelI} for further discussion.}
\label{figure_allsrc_modelI}
\end{figure*}

Qualitatively, the structure of the NGC~300 \lognlogs distribution is similar to that of the SMC \cite[][see also Figure~\ref{figure_allsrc_modelI} and next section]{Shty+Gilfanov05}, which is dominated by HMXBs due to the low stellar mass and recent elevated SFR of the SMC \citep{Antoniou+10, McSwain+05, Shty+Gilfanov05}. Two-sided Kolmogorov-Smirnoff (K-S) tests were performed to determine the probability that the differential \lognlogs distributions were drawn from the same underlying distribution; if the distributions changed significantly between observations, we would expect K-S values below a few percent. The K-S probability between ObsID 12238 and 16028 is 87\%, between ObsID 12238 and 16029 is 28\%, and between ObsID 16028 and 16029 is 2\%. We therefore find evidence that the \lognlogs distribution sometimes varies between observations.

\begin{figure}
\centering
\includegraphics[width=1\linewidth]{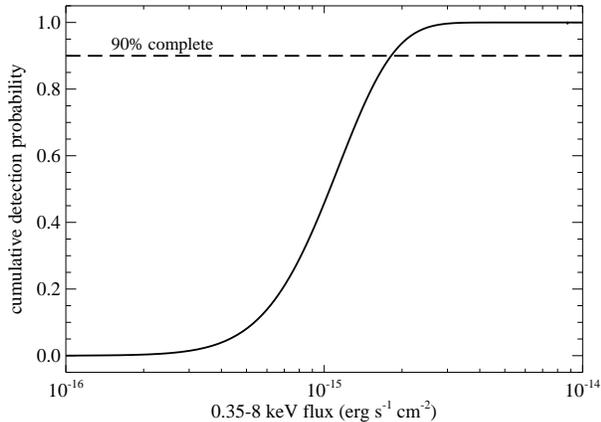}
\caption{The cumulative probability that a source with a given flux would be detected in our shallowest \Chandra exposure (ObsID 16029), calculated following the approach of \cite{Georgakakis+08}. The detection probability is folded into our Sherpa analysis of the \lognlogs distribution as an ARF; see Section~\ref{xlfs} for details.}
\label{figure_lim_flux}
\end{figure}

All fitting of the \lognlogs distributions was performed using Sherpa \citep[version 1 for CIAO 4.8;][]{Freeman+01,Doe+07} . There are two approaches to fitting the \lognlogs distributions: one can either fit the differential distribution (which may be biased by choice of binning scheme) or the cumulative distribution (which is not straightforward as the errors in each bin are correlated). In our analysis, we fit the differential distributions using the maximum-likelihood based \texttt{cstat} ($C$) statistic and the \texttt{neldermead} optimization method. Parameter uncertainties were measured using pyBLoCXs, a Markov chain Monte Carlo-based algorithm, written in Python, designed to carry out Bayesian analysis in the Sherpa environment\footnote{See \url{http://hea-www.harvard.edu/astrostat/pyblocxs/}} using the ``MetropolisMH'' sampler and 10$^4$ draws. Lower bounds were set at the 16th percentile value and upper bounds were set to the 84th value. Each free parameter was assigned a Gaussian prior centered at the best-fit value and a FWHM set by the 1.6$\sigma$ ($\sim$90\% confidence) range returned by the covariance function. There is no difference between the fit parameters that we report and those that are obtained with only the differential \lognlogs distributions, except for the size of the parameter uncertainties.

To account for incompleteness in our survey, we compute an ancillary response function (ARF) that was folded in with the \lognlogs model in Sherpa, as was done in a similar study of the Antennae \citep{Zezas+07}. Due to the low background and negligible diffuse X-ray emission from NGC~300, the ARF essentially contains the probability that a source of a given flux was detected in our survey. To calculate the ARF, we use the approach of \cite{Georgakakis+08}; we summarize the method here, and the reader is referred to their Section~4 for further details. Source extraction algorithms, such as \texttt{wavdetect}, estimate the probability that the observed number of counts within a detection cell arise from random fluctuations above the background level. Each cell contains counts from the background and possibly a source. Background images were produced using the CIAO task \texttt{flux\_image} with point sources masked, and we calculate the Poisson probability in each cell in our image that the observed number of counts could fluctuate above $L$, the minimum number of counts for a formal detection \citep[see][their equation~3]{Georgakakis+08}. The faintest source included in our survey (e.g., that made the $>$3.5$\sigma$ cut, as described in Section~\ref{section_detect}) contained $\sim$5 counts in the 0.35-8 keV band. We therefore use $L = 5$, and assume each source has a power law spectral shape with $\Gamma = 1.7$. Figure~\ref{figure_lim_flux} shows the cumulative probability that a source with a given flux would be detected in our shallowest exposure (e.g., the ARF).

We fit the three observations of the \lognlogs distribution using two physically motivated models: first, using the well-studied HMXB and LMXB XLFs (along with a background AGN component); second, we separate the X-ray sources on the basis of their temporal properties (persistent vs. variable sources, with a background AGN component).

\subsection{Model~I: AGN + HMXBs + LMXBs}\label{section_modelI}
The total X-ray luminosity from a galaxy (due to the discrete X-ray sources) is given by $L_X = \alpha$M$_* + \beta$SFR, where the coefficients $\alpha$ and $\beta$ have been measured by \cite{Lehmer+10}. HMXBs dominate the X-ray output when the SFR/M$_*$ of the host galaxy is $\gtrsim$5.9$\times10^{-11}$ yr$^{-1}$ \citep{Lehmer+10}. NGC~300 has a SFR of $\sim$0.15 \Msun yr$^{-1}$ \citep{Gogarten+10} and a stellar mass of $\sim$2$\times10^9$ \Msun \citep{Munoz+07}, which yields a SFR/M$_*$ ratio just over the HMXB-dominant threshold. Using the \cite{Lehmer+10} coefficients predicts $L_{\rm X, HMXB}\sim2.4\times10^{38}$ \lum and $L_{\rm X, LMXB}\sim1.8\times10^{38}$ \lum.

\begin{table*}[ht]
\centering
\caption{Measured Differential \lognlogs Components from the Literature}
\begin{tabular}{ccccccc} 
\hline \hline
Component	& $K^a$	& $S_{\rm b}$ (10$^{-14}$ \flux)	& $\gamma_{\rm f}$		& $\gamma_{\rm b}$	& Predicted \#$^b$	& Reference	\\
(1)			& (2)		& (3)							& (4)					& (5)				& (6)			& (7)		\\
\hline
AGN			& 395						& 1.05$\pm$0.16			& 1.55$\pm$0.18		& 2.46$\pm$0.08		& 37 (7)	& \cite{Cappelluti+09}	\\
LMXB		& (4.6$\pm$1.3)/($10^{10}$ \Msun)	& 11.7$^{+3.5}_{-4.9}$		& 1.1$^{+0.12}_{-0.13}$	& 2.4$\pm$0.5			& 4 (4)	& \cite{Lin+15}$^c$			\\
HMXB		& (1.49$\pm$0.07)$\times$SFR	& 2254$^{+1168}_{-697}$		& 1.58$\pm$0.02		& 2.73$^{+1.58}_{-0.54}$	& $<$3	& \cite{Mineo+12}		\\
\hline \hline
\label{table_physical_models}
\end{tabular}
\tablecomments{$^a$The HMXB and LMXB component normalizations are dependent upon the SFR and stellar mass of the host galaxy, respectively. The AGN normalization was not fit as a free parameter in \cite{Cappelluti+09}, but rather calculated as the value required to reproduce the number of observed AGN in their survey. $^b$The number in the parentheses gives the uncertainty in the number of object predicted. $^c$See also \cite{Kim+09}, \cite{Lehmer+14}, and \cite{Peacock+16}.}
\end{table*}

\begin{table}[ht]
 \setlength{\tabcolsep}{4pt}
\centering
\caption{Best-Fit Normalizations for Model~I}
\begin{tabular}{ccccc} 
\hline \hline
\multirow{2}{*}{Component}	& \multirow{2}{*}{Predicted \#}	& \multicolumn{3}{c}{ObsID}			\\ \cline{3-5}
						& 						& 12238		& 16028		& 16029	\\
(1)						& (2)						& (3)			& (4)			& (5)		\\
\hline
AGN			& $\sim$63-80		& 32.3$^{+4.7}_{-4.6}$	& 26.1$^{+6.7}_{-6.0}$	& 33.1$^{+6.6}_{-7.7}$	\\
LMXB		& $<$4			& $<$2.6				& $<$3.5				& $<$2.7				\\
HMXB		& $\sim$17-32		& $<$6.5				& $<$12.4				& $<$11.4				\\
\hline \hline
\label{table_ModelI_norms}
\end{tabular}
\end{table}

We first attempted to model the NGC 300 \lognlogs distribution as the sum of three components (hereafter referred to as Model~I): an AGN component, an HMXB component, and an LMXB component. All three components have had their \lognlogs distributions separately modeled by numerous authors as broken power laws with the general (differential) form

 \begin{equation}
 \frac{dN}{dS} = \begin{cases}
        K S^{-\gamma_{\rm f}}, & S < S_{\rm b}
        \\
        K S_{\rm b}^{\gamma_{\rm b}-\gamma_{\rm f}} S^{-\gamma_{\rm b}}, & S > S_{\rm b}.
        \end{cases}
 \end{equation}
 
\noindent The values of normalization constants $K$, break fluxes $S_{\rm b}$, and the faint- and bright-end slopes ($\gamma_{\rm f}$ and $\gamma_{\rm b}$, respectively) are summarized in Table~\ref{table_physical_models}. The high-luminosity break in the HMXB XLF near $\sim10^{40-41}$ \lum (e.g., $\sim$10$^{-11}$ \flux) has been observed by other authors \citep{Grimm+03,Jeltema+03} for galaxies with especially high SFRs. NGC~300 does not contain such luminous X-ray sources; the brightest X-ray source, NGC~300 X-1, has a luminosity of $\sim$4$\times10^{38}$ \lum \citep{Binder+11,Binder+15}. We therefore use a high-luminosity cut-off of 10$^{41}$ \lum for the HMXB XLF. We use the \cite{Lin+15} XLF for field LMXBs in NGC~3115 (as opposed to those found in globular clusters), which reaches a similar depth to our own observations. Other studies of the XLF of field LMXBs in early-type galaxies \citep[e.g., ][]{Kim+09,Lehmer+14,Peacock+16} and have yielded similar fit parameters. Accounting for the variance in the reported field LMXB XLFs yields an additional $\sim$30\% uncertainty in the number of LMXBS in NGC~300. Contamination from LMXBs in globular clusters is expected to be minimal, as the NGC~300 disk has been imaged multiple times by the {\it Hubble Space Telescope}, as discussed in \cite{Binder+12}.

The measured AGN source density of $\sim$480 deg$^{-2}$ \citep{Cappelluti+09} predicts $\sim$37 AGN in our survey area. We can use the HMXB and LMXB \lognlogs distributions to estimate the number of expected HMXBs and LMXBs be present in our survey. The normalizations for the HMXB and LMXB XLFs are correlated with the SFR and stellar mass of the host galaxy, respectively \citep{Mineo+12,Lin+15}; assuming a SFR of $\sim$0.15 \Msun yr$^{-1}$ and a stellar mass of $\sim$2$\times10^9$ \Msun for NGC~300 \citep{Munoz+07} yields $K_{\rm HMXB}\sim0.22$ and $K_{\rm LMXB}\sim0.92$. Integrating the differential \lognlogs distributions for these two components predicts $<$3 HMXBs and $\sim$4 LMXBs above the flux limit of our survey. These estimates suggest that $\sim$84\%, 7\%, and 9\% of X-ray sources in NGC~300 will be AGN, HMXBs, and LMXBs, respectively. The predicted number of X-ray sources ($\sim$44) is a factor of $\sim$2 lower than the observed number of X-ray sources, with the largest discrepancy likely originating in the predicted number of HMXBs \citep[see, e.g.][]{Binder+12,Williams+13}. We note that the shapes of the HMXB and LMXB XLFs have been derived using galaxies with systematically brighter X-ray point source populations than NGC~300; the other galaxy nearby galaxy with a well-studied, faint X-ray source population is the SMC, which also shows an HMXB excess. Although this excess has been attributed to a burst of recent star formation and low metallicity, low-intensity X-ray variability may be a contributing factor (see next section).

A cumulative power law distribution of the form $N(>S)\propto S^{-\gamma}$ will have a corresponding $dN/dS \propto S^{-\gamma-1}$, which is also a power law; the cumulative power law index $\gamma_{\rm c}$ is related to the differential power law index $\gamma_{\rm d}$ such that $\gamma_{\rm d}=\gamma_{\rm c}+1$. We fit the differential \lognlogs distributions using the same model (\texttt{bpl1d + bpl1d + powlaw1d} in Sherpa), with the power law indices and break fluxes frozen at the values listed in Table~\ref{table_physical_models}. Only the normalization of each component was left as a free parameter.

The best-fit normalizations for Model~I are summarized in Table~\ref{table_ModelI_norms}. The \lognlogs distribution for ObsID 16028 is shown in Figure~\ref{figure_allsrc_modelI} with both the predicted \lognlogs distribution (based on the parameters in Table~\ref{table_physical_models}) and the best-fit Model~I superimposed; fits to the other two ObsID distributions are similar. These normalizations predict that $\sim$64-80\% of X-ray sources are AGN, $\sim$2-4\% are LMXBs, and $\sim$17-32\% are HMXBs. This model does not adequately match the observed \lognlogs distributions; typically, $\chi^2$/dof $\sim270/71$, while the $Q$-value (the probability that one would observe the reduced statistic value, or a larger value, if the assumed model is true) returned $\sim$0 for all three observations. We therefore consider a different model for the shape of the \lognlogs distribution that can be tested with our multiple exposures of NGC~300: one in which sources are separated by their temporal variability properties.

\subsection{Model~II: AGN + Persistent XRBs + Variable XRBs}\label{section_modelII}
Given the detection of variability in the shape of the \lognlogs distribution, we next consider whether the NGC~300 X-ray point source population can be characterized by the variability properties of the XRBs instead of by the mass of their companion donor star. Both HMXB and LMXB systems exhibit X-ray variability, and both types of systems accrete via the same basic mechanisms. Systems with persistently high X-ray luminosities (above $10^{35}$ \lum) are produced when the radius of the donor star fills its Roche lobe, and mass transfer to the compact object becomes dominated by the tidal stream between the two components (e.g., Roche lobe overflow, RLOF). Although these sources can achieve luminosities up to $\sim$10$^{40}$ \lum, lower luminosities of a few $10^{36}$ \lum are frequently observed in Galactic sources \citep[e.g., Vela X-1; see][and references therein]{Walter+15} and in SMC pulsar-HMXBs \citep{Laycock+10}. However, most XRBs in the Milky Way and SMC produce very low X-ray luminosities ($\sim$10$^{33}$ \lum). Instead of undergoing RLOF, the compact objects in these systems capture only a small fraction of their companion star's wind and, therefore, have very low accretion rates \citep{Bondi+44,Davidson+73,Lamers+76,Liu+06,Laycock+05}. There is some evidence that the XLF of Galactic wind-fed systems may follow a broken power law distribution, with sources below $\sim2.5\times10^{36}$ \lum following a flatter power law index \citep[differential $\gamma\sim1.4$, at $\sim$2$\sigma$ significance;][]{Lutovinov+13}.

\begin{figure*}
\centering
\includegraphics[width=1\linewidth]{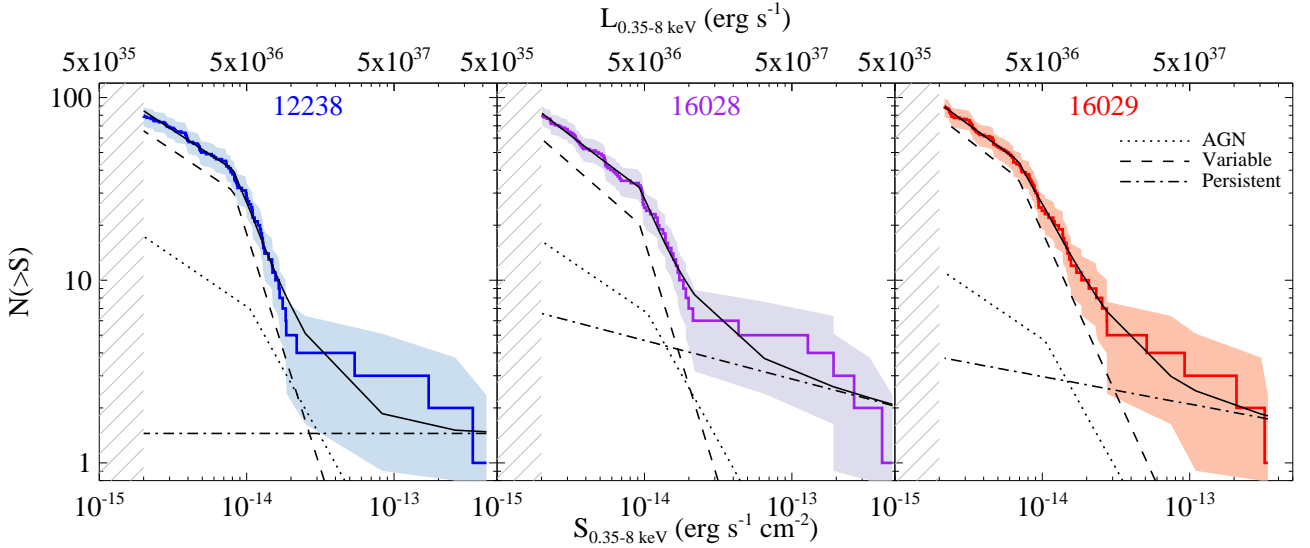} 
\caption{The cumulative \lognlogs distributions derived for each ObsID. The best-fit model is superimposed (solid black line). The AGN, variable, and persistent components are shown by the dotted, dashed, and dot-dashed lines, respectively. The gray-lined region indicates fluxes below our 90\% completeness limit; only sources above this limit are shown.}
\label{figure_allsrc_modelII}
\end{figure*}

\begin{table*}[ht]
\setlength{\tabcolsep}{4pt}
\centering
\caption{Best-Fit \lognlogs Distribution Parameters}
\begin{tabular}{ccccccccccccc} 
\hline \hline
\multirow{2}{*}{Obs ID}	& \multirow{2}{*}{$K_{\rm AGN}$}	& \multicolumn{2}{c}{Persistent Component}	&& \multicolumn{4}{c}{Variable Component}	& \multirow{2}{*}{$\chi^2$/dof$^b$} & \multicolumn{3}{c}{Predicted \# of Sources}	\\ \cline{3-4} \cline{6-9} \cline{11-13}
					& 		& $K$	& $\gamma$			&& $K$	& $S_{\rm b}^a$	& $\gamma_{\rm f}$		& $\gamma_{\rm b}$		&	& AGN	& Persistent	& Variable	\\
(1)					& (2)		& (3)		& (4)					&& (5)	& (6)				& (7)					& (8)					& (9)	& (10)	& (11)		& (12)	\\
\hline
12238	& 8.5$^{+4.2}_{-4.3}$	& 0.7$\pm$0.4 			& 1.3$\pm$0.1	&& 14.6$^{+2.1}_{-2.2}$	& 0.9$\pm$0.1	& 1.6$\pm$0.2			& 4.9$^{+1.3}_{-1.2}$	& 59/69	& 21$^{+10}_{-11}$	& 4$\pm$2	& 42$\pm$14	\\
16028	& 8.4$^{+4.5}_{-4.7}$	& 1.4$^{+0.5}_{-0.6}$	& $<$1.2		&& 12.0$^{+2.1}_{-2.0}$	& 0.9$\pm$0.1	& 1.8$^{+0.2}_{-0.3}$	& 2.8$^{+1.8}_{-0.8}$	& 48/67	& 20$\pm$11		& 9$^{+3}_{-4}$	& 44$^{+30}_{-17}$	\\
16029	& 6.7$\pm$4.4			& 0.9$\pm$0.4			& $<$1.2		&& 17.6$^{+2.7}_{-2.0}$	& 0.7$\pm$0.1	& 1.7$\pm$0.2			& 3.5$\pm$1.1			& 43/76	& 16$\pm$11		& 6$\pm$3	& 55$^{+22}_{-21}$	\\
\hline \hline
\label{table_fits_allsrc}
\end{tabular}
\tablecomments{$^a$Break flux is given in units of 10$^{-14}$ \flux. $^b$Degrees of freedom.}
\end{table*}

High X-ray luminosities in non-RLOF HMXBs systems are produced in outbursts, which can be classified into Types~I and II. Type~I outbursts, which can reach luminosities of $\sim$10$^{37}$ \lum, typically corresponding to the periastron passage of a NS in an eccentric orbit about its companion. As the NS enters the denser portions of the stellar wind close to periastron, the accretion rate increases and produces a predictable increase in the observed X-ray luminosity. Type~II outbursts are somewhat fainter, $\sim10^{36-37}$ \lum, and can last from minutes to several orbital periods. These types of outbursts are frequently observed in supergiant HMXBs \citep[see][and references therein]{Reig08,Ducci+14}, and the so-called supergiant fast X-ray transient \citep[SFXT,][]{Negueruela+08} events are shorter X-ray flares that occur as part of much longer outburst events, which can last several days \citep{Sidoli+08,Romano+11,Romano+14}. The mechanism by which Type~II outbursts and SFXT flares are produced is not certain, although eruptions from the donor star or changes in the stellar wind properties may contribute to the observed X-ray variability \citep[][and references therein]{Ducci+14}. LMXBs also exhibit strong X-ray variability, which is frequently accompanied by spectral changes that are driven by the structure of the inner accretion disk \citep{Lewin+97,Maccarone03,Asai+12}.

\begin{figure}
\centering
\includegraphics[width=1\linewidth]{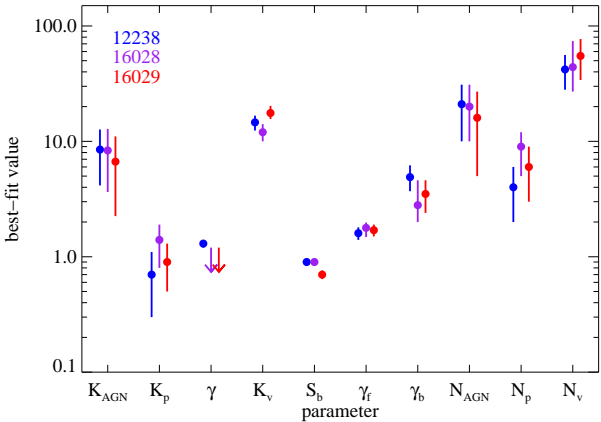} 
\caption{Best-fit parameters, and the predicted number of sources, to the \lognlogs distributions for each ObsID.}
\label{figure_allsrc_compare}
\end{figure}

We modeled the NGC~300 \lognlogs distribution as the sum of an AGN component (as in the previous section), a persistent XRB component (assumed to be a simple power law), and a variable XRB component. The variable component was modeled as a broken power law, as we anticipate a sharp decline in the number of observed sources above the typical outburst luminosity. The average AGN variability amplitude is $\sim$25-30\% \citep{Soldi+14}, which is comparable to our flux uncertainties, particularly at the faint end; thus, we assume the background AGN contribution to our observed X-ray point sources is constant. The best-fit \lognlogs models are shown in Figure~\ref{figure_allsrc_modelII}, and a visual comparison of the best-fit parameters is shown in Figure~\ref{figure_allsrc_compare}.

Model~II is a significantly better description than Model~I of the \lognlogs distributions in all three observations; an F-test indicates that the improvement is significant at the $\sim$10$\sigma$ level. We note that the power law index and the predicted number of persistent sources is similar to the expected LMXB population. The power law index for the persistent component is $\lesssim$1.3, compared to an expected LMXB faint-end $\gamma_{\rm c}\sim1-1.2$. A similar result is found for variable sources and HMXBs: the faint-end of the variable broken power law model has a $\gamma_{\rm f}\sim1.6-1.8$, similar to $\gamma_{\rm c}\sim1.6$ for the HMXB XLF. This result suggests that LMXBs may be, as a population, more persistent X-ray emitters (at least over the flux range sampled by our survey), while HMXBs show a greater degree of variability.

	\subsection{Persistent vs. Variable Sources}
The NGC~300 \lognlogs distribution is best described as a combination of persistently bright XRBs and AGN and variable XRBs. To examine the contribution of these two populations to the overall \lognlogs distribution, we divided our sample into two categories: ``persistent'' sources that were detected in all three observations and did not show a flux change of more than a factor of two (within the flux uncertainties), and ``variable'' sources that either were detected in all observations but showed more than a factor of two change in flux, or were not detected in at least one observation but had a flux more than a factor of two above the 90\% limiting flux in a different observation. We found 31 sources met our ``persistent'' criteria, and 41 sources were classified as ``variable.'' This is roughly consistent with the predicted number of persistent ($\sim$20--27) and variable ($\sim$33--49) sources from the previous section, as AGN are expected to meet this definition of ``persistent.'' The remaining 13 sources in the common area of our observations had unknown or ambiguous variability properties (e.g., they did not exceed a factor of two above the 90\% limiting flux in one or two observation in which they were not detected) and so were not used in this analysis. A discussion of the \lognlogs distribution properties as a function of {\it degree} of variability is presented in the next subsection.

The 0.35-8 keV \lognlogs distributions were calculated in each epoch for the persistent and variable sources. There are three ways these distributions may be compared to one another: the persistent source distributions across all three epochs, the variable source distributions across all three epochs, and the persistent vs. variable source distributions within a single epoch. There is marginal evidence that the persistent source \lognlogs distributions vary across all three epochs; the minimum KS probability was 10\% between ObsID 16028 and 16029. However, the variable source logN-logS distributions show significant differences between observations: the KS probability is $\lesssim$1.4\% for all three combinations of epochs. Table~\ref{table_ks_obs1} summarizes the KS probabilities for the \lognlogs distributions for all sources, persistent sources, and variable sources between observations. The probability that the persistent source distribution and the variable source distribution were drawn from the same underlying distribution {\it within a single observation} was also calculated and found to be $<0.05\%$ for all three observations. We can therefore say with confidence that the persistent X-ray sources in NGC~300 have fundamentally different population properties than the variable sources.

\begin{table}[ht]
\centering
\caption{K-S Test Probabilities: By ObsID}
\begin{tabular}{c|cccccccc} 
\hline \hline
\multirow{2}{*}{ObsID}	& \multicolumn{2}{c}{All Sources}	&& \multicolumn{2}{c}{Persistent}	&& \multicolumn{2}{c}{All Variable} \\ \cline{2-3} \cline{5-6} \cline{8-9}
					& 16028	& 16029				&& 16028	& 16029				&& 16028	& 16029			\\ 
\hline
12238	& 0.87	& 0.28	&& 0.96	& 0.55	&& 0.0016		& 0.0039	\\
16028	& 1		& 0.02	&& 1		& 0.10	&& 1			& 0.0135	\\
\hline \hline
\end{tabular}
\tablecomments{Two-sided K-S probabilities that two \lognlogs distributions are drawn from the same distribution.}
\label{table_ks_obs1}
\end{table}

\begin{figure*}[ht]
\centering
\begin{tabular}{c}
\includegraphics[width=1\linewidth,clip=true,trim=0cm 1.3cm 0cm 0.2cm]{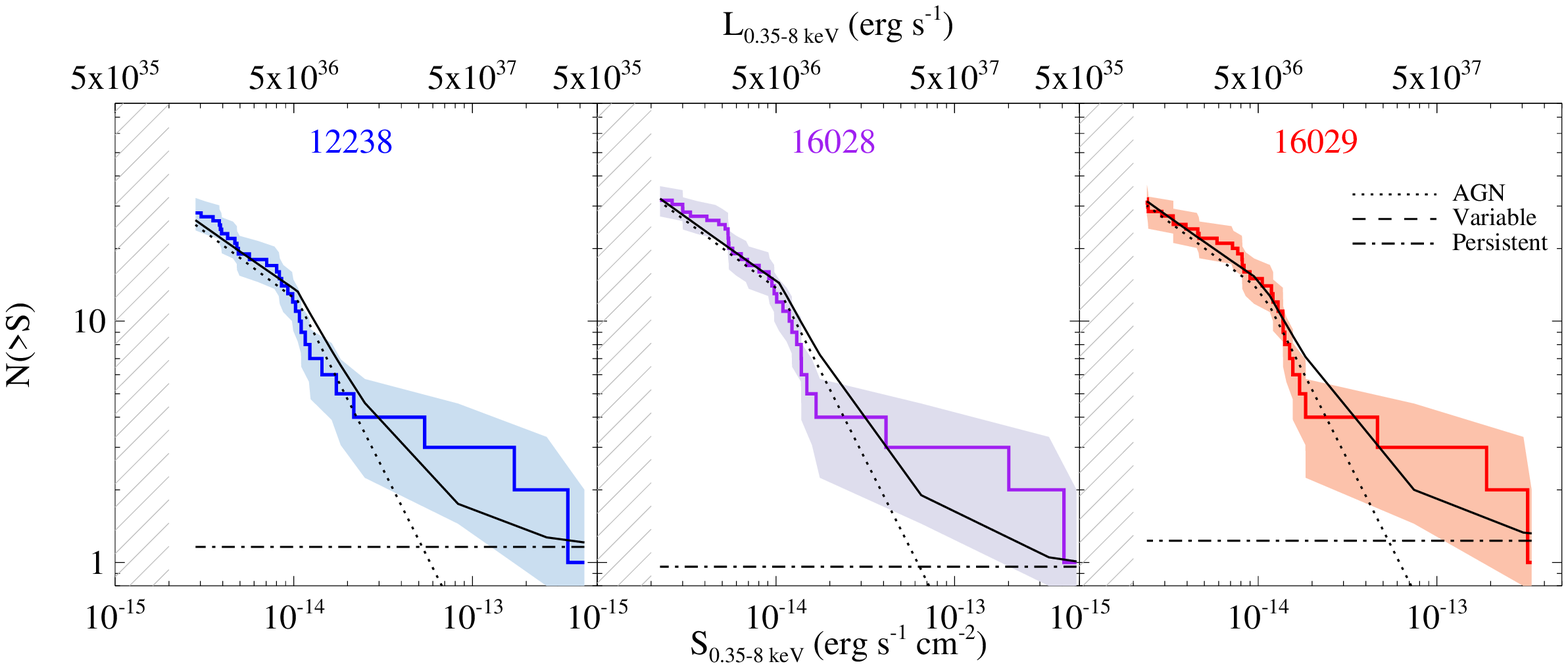} \\
\includegraphics[width=1\linewidth,clip=true,trim=0cm 0cm 0cm 2.0cm]{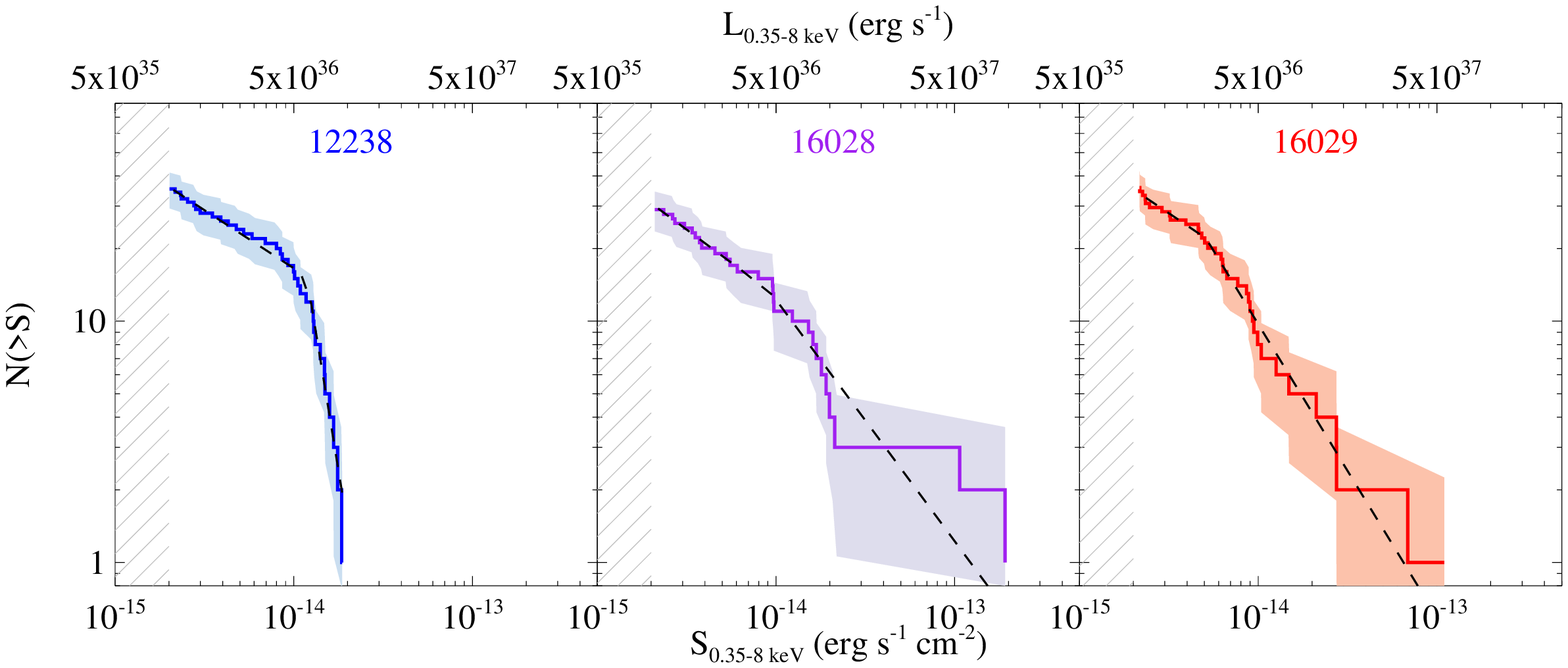} \\
\end{tabular}
\caption{The cumulative \lognlogs distributions of persistent sources (top) and variable sources (bottom). The AGN, variable, and persistent components are shown by the dotted, dashed, and dot-dashed lines, respectively. The persistent sources were fit with AGN and persistent components (the solid black line shows the sum of the components), while the variable sources were fit only with a variable component. ObsID 16028 shows the greatest similarity in the \lognlogs distributions between persistent and variable sources. The gray-lined region indicates fluxes below our 90\% completeness limit; only sources above this limit are shown.}
\label{figure_fit_per_var}
\end{figure*}

Both distributions were initially fit as a three-component model: the AGN component, a broken power law (the variable component), and a simple power law (the persistent component), as in the previous section. When this model was applied to the persistent source distribution, the normalization of the variable component was consistent with zero. Likewise, the normalizations of the AGN and persistent components were both consistent with zero when we attempted to fit the variable sources with a three-component model. That the AGN component normalization was consistent with zero is not surprising. Studies of AGN variability as a function of flux \citep[e.g., ][]{Paolillo+04} suggest that only $\sim$10\% of AGN with $<$100 net counts (i.e., the majority of our sample) would exhibit X-ray flux variations significant enough to be considered ``variable'' in our study. We therefore removed these components and re-fit the distributions with the simplified versions of the model. The best-fit parameters found in both cases were nearly identical, but the simplified models yielded smaller uncertainties on the fit parameters; for example, typical uncertainties for the persistent source fit parameters were $\sim$19\% when fitting three components, but $\sim$11\% when fitting with two. F-tests between the simplified model and those with additional components yielded probabilities of $\sim$50--99\%, indicating the additional components did not improve the quality of the fits. The results are shown in Figure~\ref{figure_fit_per_var}, and the best-fit parameters are summarized in Table~\ref{table_fits_per_var}.

\begin{table*}[ht]
\setlength{\tabcolsep}{4pt}
\centering
\caption{Best-Fit Persistent \& Variable Source \lognlogs Distribution Parameters}
\begin{tabular}{ccccccccccccccc}
\hline \hline
Source	& \multirow{2}{*}{Obs ID}	& \multirow{2}{*}{$K_{\rm AGN}$}	& \multicolumn{2}{c}{Persistent Component}	&& \multicolumn{4}{c}{Variable Component}	& \multirow{2}{*}{$\chi^2$/dof$^b$}	& \multicolumn{3}{c}{Predicted \# of Sources} \\ \cline{4-5} \cline{7-10} \cline{12-14}
Type		&			& 					& $K$	& $\gamma$			&& $K$	& $S_{\rm b}^a$	& $\gamma_{\rm f}$		& $\gamma_{\rm b}$	&		& AGN	& Persistent	& Variable	\\
(1)		& (2)			& (3)					& (4)		& (5)					&& (6)	& (7)				& (8)					& (9)				& (10)	& (11)	& (12)		& (13) \\
\hline
\multirow{3}{*}{Persistent}	& 12238	& 12.5$^{+1.8}_{-1.0}$	& 0.4$\pm$0.2	& $<$1.4	&& ...	& ...	& ...	& ...	& 35/23	& 30$^{+5}_{-2}$	& 2$\pm$1	& ...	\\
					& 16028	& 13.7$^{+1.3}_{-0.9}$	& 0.4$\pm$0.2	& $<$1.5	&& ...	& ...	& ...	& ...	& 43/26	& 33$^{+4}_{-2}$	& 2$\pm$1	& ...	\\
					& 16029	& 13.7$^{+1.0}_{-1.3}$	& 0.4$\pm$0.2	& $<$1.5	&& ...	& ...	& ...	& ...	& 27/26	& 33$\pm$3		& 2$\pm$1	& ...	\\
\hline
\multirow{3}{*}{Variable}	& 12238	& ... & ...		& ...		&& 11.5$^{+1.5}_{-1.6}$	& 1.2$\pm$0.1			& 1.4$\pm$0.2			& 5.7$^{+1.6}_{-2.8}$	& 13/30	& ...	& ...	& 30$^{+11}_{-16}$	\\
					& 16028	& ... & ...		& ...		&& 7.3$^{+2.7}_{-2.2}$	& 0.9$^{+0.2}_{-0.5}$	& 1.5$\pm$0.3			& 2.1$^{+0.6}_{-0.5}$	& 16/23	& ...	& ...	& 25$^{+14}_{-18}$	\\
					& 16029	& ... & ...		& ...		&& 13.8$^{+5.3}_{-3.5}$	& 0.5$\pm$0.3			& 1.3$^{+0.5}_{-0.2}$	& 2.3$^{+1.2}_{-0.7}$	& 17/29	& ...	& ...	& 30$^{+29}_{-22}$	\\
\hline \hline
\label{table_fits_per_var}
\end{tabular}
\tablecomments{$^a$Break flux is given in units of 10$^{-14}$ \flux. $^b$Degrees of freedom.}
\end{table*}

The persistent source distribution is dominated by the AGN component, which predicts $\sim$33 AGN within the common area of our survey. This is roughly consistent with the number we expect ($\sim$37) given the \cite{Cappelluti+09} AGN source density. The variable source distribution, on the other hand, is unlikely to have significant contamination by AGN. The low number of persistent sources intrinsic to NGC~300 ($\sim$2) and the power law slope ($<1.4$) is similar to the field LMXB XLF \citep{Lin+15}. This component is needed to explain the bright-end of the observed \lognlogs distributions, which is dominated by the bright source NGC~300 X-1 \citep[][and references therein]{Binder+11,Binder+15}. Although previously thought to be a Wolf Rayet + black hole HMXB, recent observations by \cite{Binder+15} have suggested that the donor star may be significantly less massive than previously believed, making X-1 a persistently bright black hole-LMXB. 

The best-fit parameters for the persistent source distributions are also similar to those that were found for the persistent component in Section~\ref{section_modelII}. The component normalization, break fluxes, and bright end power law indices for the variable sources exhibit more significant variation between exposures. The break fluxes ObsID 12238 and 16028 are similar, $\sim9\times10^{-15}$ \flux vs. $\sim1.2\times10^{-14}$ \flux, respectively, but quite different from the $\sim5\times10^{-15}$ \flux found in ObsID 16029. Despite the differences in break flux, however, ObsID 16028 and 16029 have very similar bright-end power law indices (2.1 and 2.3, respectively). This difference in the bright-end slope from ObsID 12238 ($\gamma_{\rm b}\sim$5.7) is a consequence of a small number of bright sources observed in ObsIDs 16028 and 16029. Interestingly, the most stable component of the variable source distribution is the faint-end power law index, which is $\sim$1.4 in all three exposures. The break fluxes correspond to luminosities of $\sim$(2--67)$\times10^{36}$ \lum at the distance of NGC~300. These parameters are consistent with the low-luminosity, wind-fed Galactic HMXB XLF presented in \cite{Lutovinov+13}, whose definition of ``persistent'' refers to a lack of rapid X-ray variability on the order of the exposure time, whereas we are considering variability over months and years. Although the number of variable sources is consistent with the expected number of HMXBs from Table~\ref{table_ModelI_norms}, we cannot rule out the possibility that variable LMXBs are contributing to the observed \lognlogs distributions.

We use the best-fit \lognlogs distributions to estimate the total X-ray luminosity produced by variable sources and persistent sources that are intrinsic to NGC~300. To do this, we randomly select the number of sources described by each component, drawn from the range of expected number of sources in Table~\ref{table_fits_per_var} (e.g., $\sim$30 sources). We then populate the best-fit power law (or broken power law) with this number of sources and calculate the resulting luminosity. We repeat this process 10$^4$ times for each distribution to estimate the range in X-ray luminosity our models predict from NGC~300. The persistent sources are expected to produce a luminosity of (1.8$^{+12.0}_{-1.5}$)$\times10^{37}$ \lum, while the variable sources collectively produce (2.5$^{+10.7}_{-2.2}$)$\times10^{38}$ \lum. The large uncertainties in the persistent source luminosity is due to the small number of X-ray sources intrinsic to NGC~300 compared to AGN (e.g., $\sim$2 persistent XRBs are expected, compared to $\sim$30 AGN), while the uncertainties in the variable source luminosity is primarily driven by the large differences in observed $\gamma_{\rm b}$. Within the uncertainties, however, these luminosities are consistent with the predicted LMXB and HMXB luminosities.

	\subsection{Variability Subclasses}
We next considered whether the degree of individual source variability influenced the shape of the \lognlogs distributions. The variable sources were separated into three subclasses: ``low-level'' variable sources showed flux variations more than a factor of two but less than a factor of four, and ``intermediate'' variable sources exhibited flux variations of more than a factor of four but less than a factor of ten. ``Transient'' sources were not detected in at least one exposure, but had a flux more than an order of magnitude above the 90\% limiting flux in at least one observation {\it or} exhibited a change in flux greater than an order of magnitude.  Table~\ref{table_variable_classes} summarizes the definitions of our variable source classification scheme.

\begin{table}[ht]
\centering
\caption{Categories of X-ray Variability}
\begin{tabular}{ccc} 
\hline \hline
Class		& Definition	& \# Sources	\\
(1)			& (2)			& (3)	\\
\hline
persistent				& $S_{\rm max}/S_{\rm min}\leq 2$		& 31	\\
low-level variable		& $2 < S_{\rm max}/S_{\rm min} \leq 4$	& 18	\\
intermediate variable		& $4 < S_{\rm max}/S_{\rm min} < 10$	& 12	\\
\multirow{2}{*}{transient}	& $S_{\rm max}/S_{\rm min} > 10$;		& \multirow{2}{*}{11}	\\
					& at least one non-detection			& 	\\
\hline \hline
\label{table_variable_classes}
\end{tabular}
\tablecomments{$S_{\rm max}$ and $S_{\rm min}$ are the maximum and minimum observed fluxes.}
\end{table}

\begin{figure*}
\centering
\begin{tabular}{c}
	\begin{overpic}[width=1\linewidth,clip=true,trim=0cm 1.3cm 0cm 0cm]{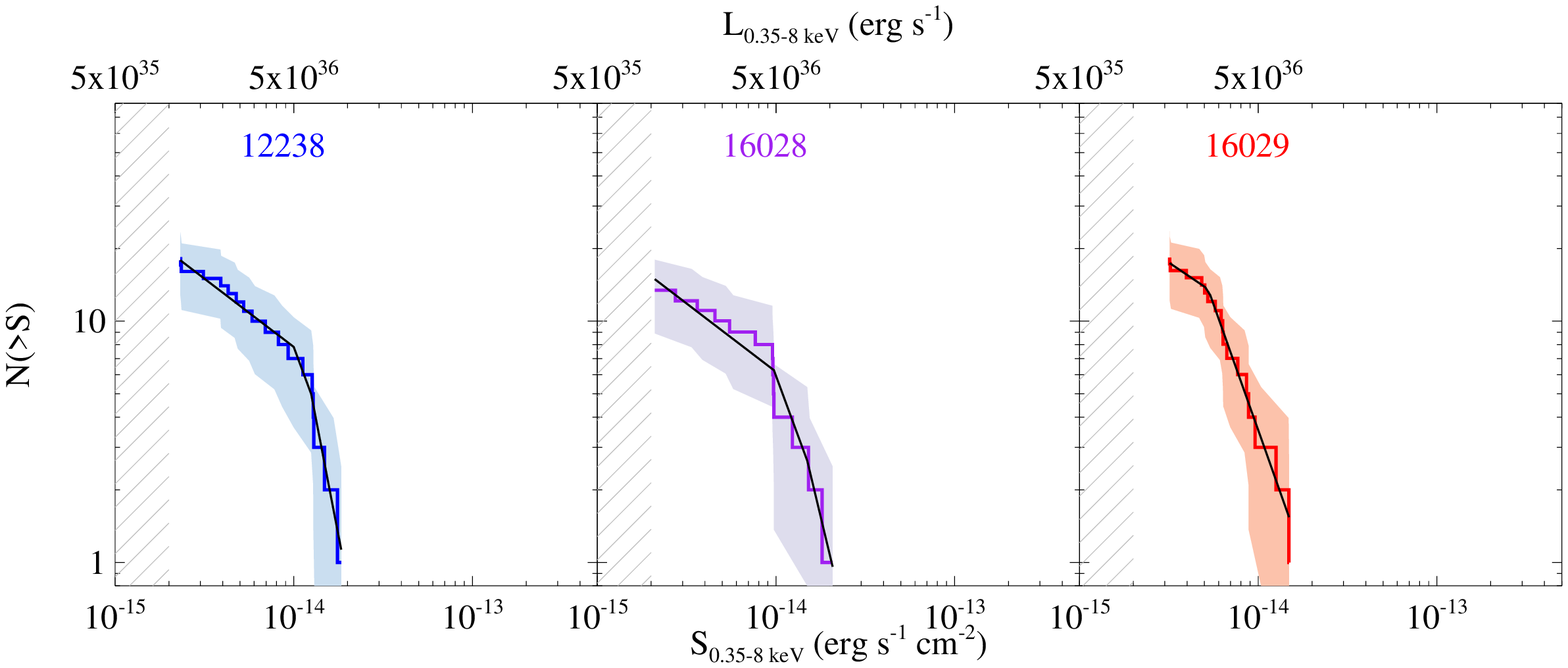}
 		\put (10,23){\large low level}
	\end{overpic} \\
	\begin{overpic}[width=1\linewidth,clip=true,trim=0cm 1.3cm 0cm 1.9cm]{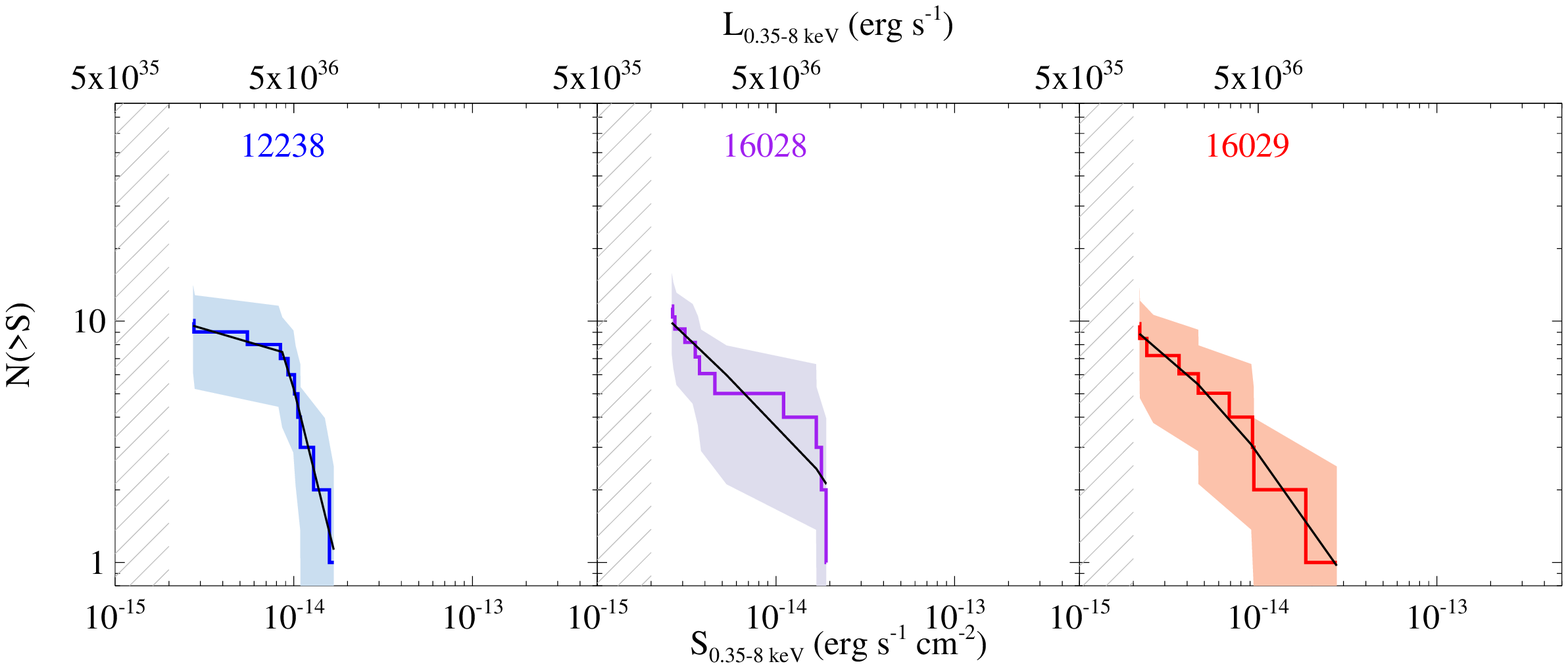}
 		\put (10,23){\large intermediate}
	\end{overpic} \\
	\begin{overpic}[width=1\linewidth,clip=true,trim=0cm 0cm 0cm 1.9cm]{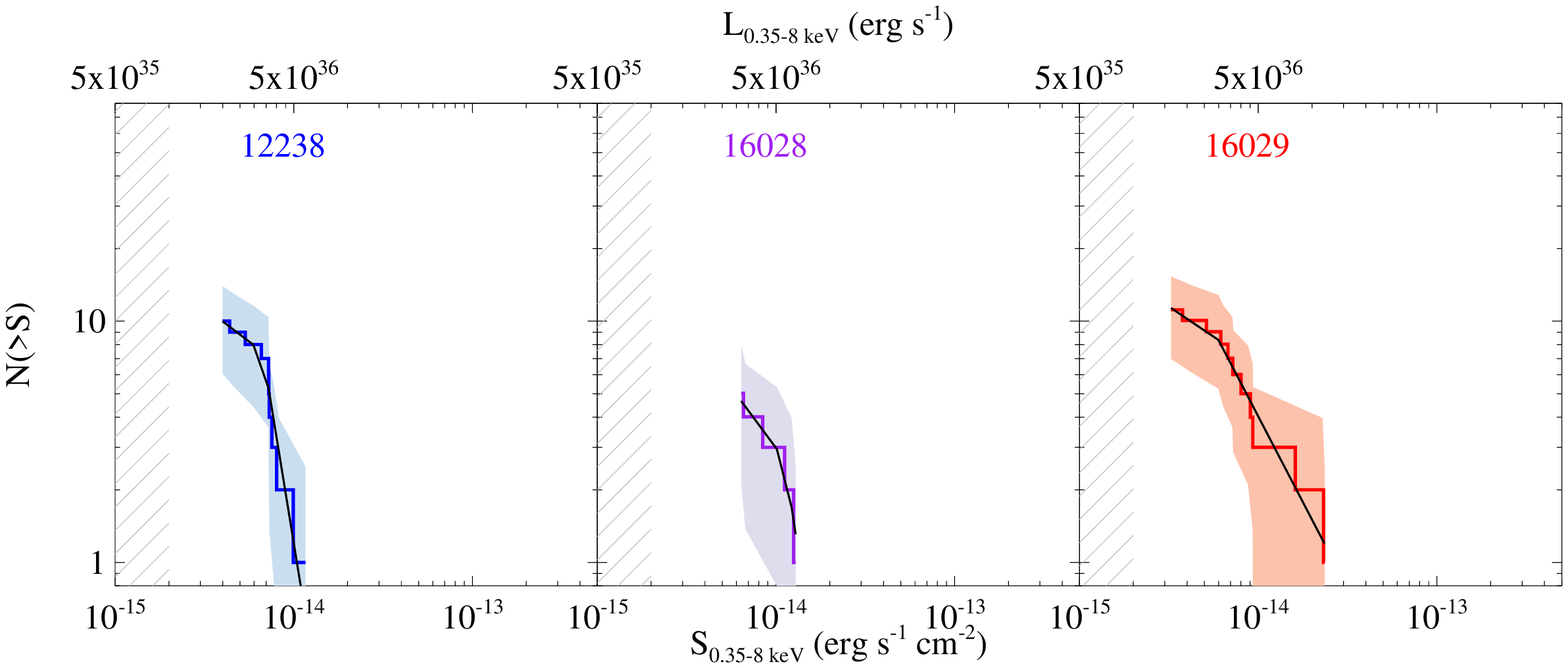}
 		\put (10,27){\large transient}
	\end{overpic} \\
\end{tabular}
\caption{The cumulative \lognlogs distributions for each ObsID for low-level variable sources (top row), intermediate variables (middle row), and transient sources (bottom row). The best-fit model is shown by the solid black line. The gray-lined region indicates fluxes below our 90\% completeness limit; only sources above this limit are shown.}
\label{figure_fit_variable}
\end{figure*}

\begin{figure}
\centering
\includegraphics[width=1\linewidth]{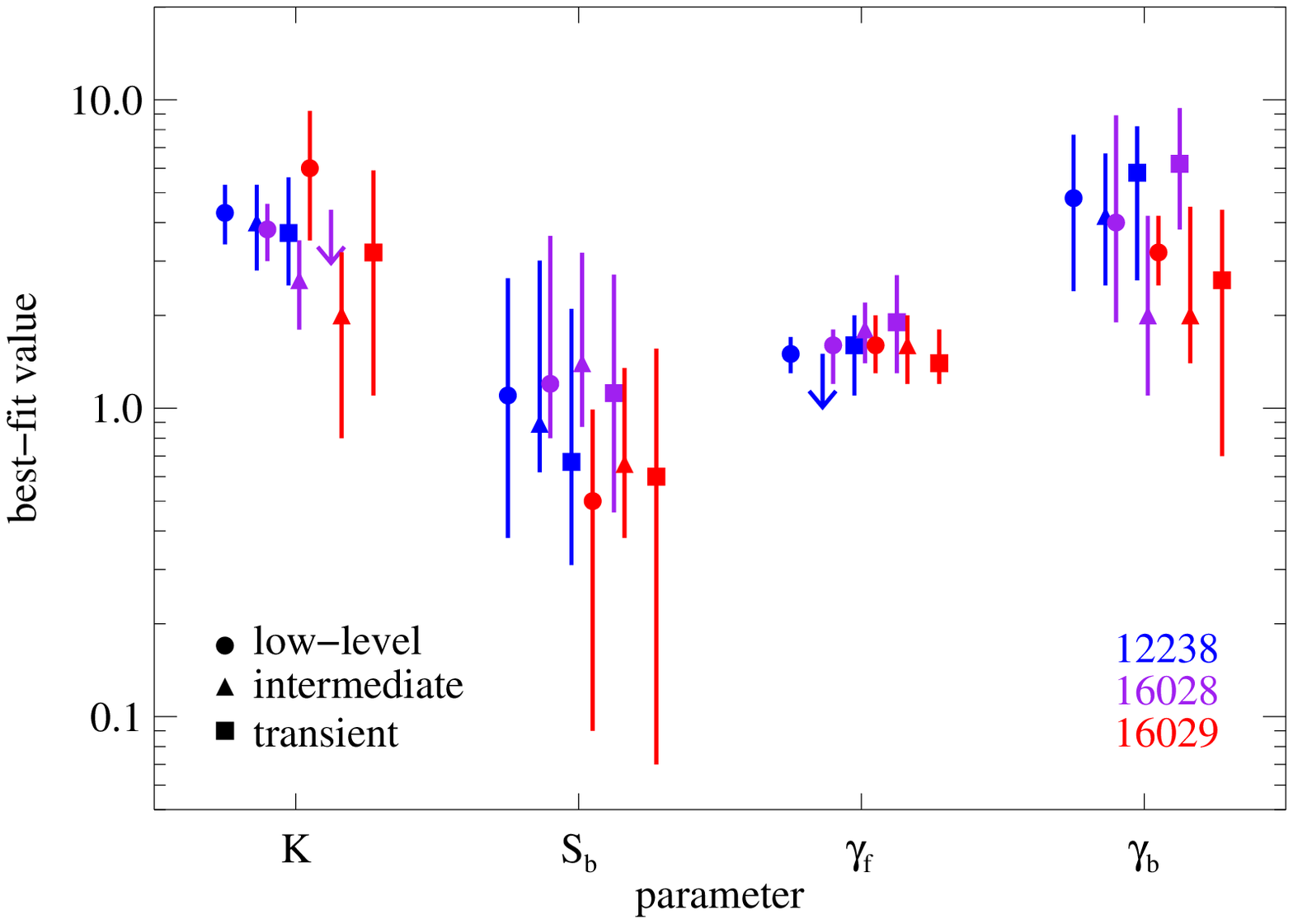}
\caption{Best-fit parameters obtained for our fits to the \lognlogs distributions for each ObsID and each variability category.}
\label{figure_variable_compare}
\end{figure}

The \lognlogs distributions were computed for each category in each observing epoch (using the source flux measured in that epoch), and a broken power law was fit to the resulting distributions. The results are shown in Figure~\ref{figure_fit_variable}. The best-fit parameters are summarized in Table~\ref{table_fit_variable} and shown in Figure~\ref{figure_variable_compare}. Although the uncertainties are large due to the small number of sources used in the fit, nearly all the \lognlogs distributions show faint-end power law indices consistent with $\sim$1.6. The typical break flux range of $\sim$(0.5--1.5)$\times10^{-14}$ \flux corresponds to a luminosity range of $\sim$(2.4--7.3)$\times10^{36}$ \lum at the distance of NGC~300. 

\begin{table*}[ht]
\centering
\caption{\lognlogs Distribution Fits for Variable Source Subclasses}
\begin{tabular}{ccccccc} 
\hline \hline
Category	& Obs ID	& $K$	& $S_{\rm b}^a$	& $\gamma_{\rm f}$	& $\gamma_{\rm b}$	& $\chi^2$/dof$^b$	 \\
(1)		& (2)		& (3)		& (4)				& (5)				& (6)				& (7)			\\
\hline
\multirow{3}{*}{low-level}		& 12238	& 4.3$^{+1.0}_{-0.9}$	& 1.1$^{+1.5}_{-0.7}$	& 1.5$\pm$0.2			& 4.8$^{+2.9}_{-2.4}$	& 5/13	\\
						& 16028	& 3.8$\pm$0.8			& 1.2$^{+2.4}_{-0.4}$	& 1.6$^{+0.2}_{-0.4}$	& 4.0$^{+4.9}_{-2.1}$	& 7/8		\\
						& 16029	& 6.0$^{+3.2}_{-2.5}$	& 0.5$^{+0.5}_{-0.4}$	& 1.6$^{+0.4}_{-0.3}$	& 3.2$^{+1.0}_{-0.7}$	& 4/13	\\
\hline
\multirow{3}{*}{intermediate}	& 12238	& 4.0$^{+1.3}_{-1.2}$	& 0.9$^{+2.1}_{-0.3}$	& $<$1.5			& 4.2$^{+2.5}_{-1.7}$	& 2/5		\\
						& 16028	& 2.6$^{+0.9}_{-0.8}$	& 1.4$^{+1.8}_{-0.5}$	& 1.8$\pm$0.4		& 2.0$^{+2.2}_{-0.9}$	& 6/6		\\
						& 16029	& 2.0$\pm$1.2			& 0.7$^{+0.7}_{-0.3}$	& 1.6$\pm$0.4		& 2.0$^{+2.5}_{-0.6}$	& 2/4		\\
\hline
\multirow{3}{*}{transient}		& 12238	& 3.7$^{+1.9}_{-1.2}$	& 0.7$^{+1.4}_{-0.4}$	& 1.6$^{+0.4}_{-0.5}$	& 5.8$^{+2.4}_{-3.7}$	& 3/5		\\
						& 16028	& $<$4.4				& 1.1$^{+1.6}_{-0.7}$	& 1.9$^{+0.8}_{-0.6}$	& 6.2$^{+3.2}_{-2.4}$	& 1/1		\\
						& 16029	& 3.2$^{+2.7}_{-2.1}$	& 0.6$^{+1.0}_{-0.5}$	& 1.4$^{+0.4}_{-0.2}$	& 2.6$^{+1.8}_{-1.9}$	& 2/6		\\
\hline \hline
\label{table_fit_variable}
\end{tabular}
\tablecomments{$^a$Break flux is given in units of 10$^{-14}$ \flux. $^b$Degrees of freedom.}
\end{table*}

Two-sided K-S tests of the \lognlogs distributions between observations did not reveal any evidence for differences within the variability subclasses -- the low-level variable sources in ObsID 12238 look, statistically, like the low-level variable sources in ObsID 16029. Table~\ref{table_ks_obs2} provides the K-S test results for all ObsID combinations. We next compared the different variability categories within a single observation (i.e., we tested if the persistent sources in ObsID 12238 differed significantly from the transient sources in the same observation). The results are summarized in Table~\ref{table_ks_var}. There is significant evidence that the persistent X-ray sources are different from all categories of variable sources. However, there is no evidence that the different categories of variable sources differ from one another, which suggests that all types of variable X-ray sources are part of the same underlying population.

\begin{table}[ht]
\centering
\caption{Variable Distribution K-S Test Probabilities: By ObsID}
\begin{tabular}{c|cccccccc} 
\hline \hline
\multirow{2}{*}{ObsID}	& \multicolumn{2}{c}{Low-Level}	&& \multicolumn{2}{c}{Intermediate}	&& \multicolumn{2}{c}{Transient}	\\  \cline{2-3} \cline{5-6} \cline{8-9}
					& 16028	& 16029				&& 16028	& 16029				&& 16028	& 16029			\\ 
\hline
12238	& 0.99	& 0.92	&& 0.99	& 0.99	&& 0.99	& 0.22	\\
16028	& 1		& 0.68	&& 1		& 0.70	&& 1		& 0.12	\\
\hline \hline
\end{tabular}
\tablecomments{Two-sided K-S probabilities that two \lognlogs distributions from the same variability class are drawn from the same parent distribution in different observations.}
\label{table_ks_obs2}
\end{table}

\begin{table}[ht]
\centering
\caption{K-S Test Probabilities: Within Each ObsID}
\begin{tabular}{c|ccc} 
\hline \hline
\multirow{2}{*}{Class}	& low level	& intermediate	& transient	\\ \cline{2-4}
		& \multicolumn{3}{c}{12238}	\\
\hline
persistent		& 1.9$\times10^{-5}$	& 2.8$\times10^{-5}$	& 0.0025	\\
low level		& 1				& 0.99			& 0.36	\\
intermediate	& ...				& 1				& 0.62	\\
\hline
		& \multicolumn{3}{c}{16028}	\\
\hline
persistent		& 2.2$\times10^{-6}$	& 2.9$\times10^{-5}$	& 0.0026		\\
low level		& 1				& 0.99			& 0.14		\\
intermediate	& ...				& 1				& 0.30		\\
\hline
		& \multicolumn{3}{c}{16029}	\\
\hline
persistent		& 0.0004	& 2.9$\times10^{-5}$	& 3.8$\times10^{-6}$	\\
low level		& 1		& 0.44			& 0.53		\\
intermediate	& ...		& 1				& 0.99		\\
\hline \hline
\end{tabular}
\tablecomments{Two-sided K-S probabilities that two \lognlogs distributions in one observation are drawn from the same parent distribution as a different variability class distribution.}
\label{table_ks_var}
\end{table}

\section{Discussion}\label{discussion}
	\subsection{Modeling the XLF of Variable Sources}
XRBs are intrinsically variable objects, and yet the resulting \lognlogs distributions constructed for an entire population of XRBs appear to follow a similar shape. We next considered whether the X-ray variability properties of these sources -- such as the peak flux during an X-ray outburst or the frequency of the outbursts -- could explain the shape of the observed \lognlogs distributions. The high-energy light curves of variable XRBs have been studied extensively using {\it RXTE}, {\it BeppoSAX}, {\it Integral}, {\it Suzaku}, and {\it Swift}, especially for Galactic sources \citep[see, e.g.][]{Reig+13,Reig11,Stroh+13} and in the SMC \citep{Laycock+05}. Generally, the shapes of these light curves fall into one of two generic classes: a smooth increase and subsequent decrease in the observed flux that broadly resembles a Gaussian curve, or a fast rise in X-ray flux followed by an exponential decay \citep{Reig11,Reig+13}. We refer to these as ``Gaussian'' and fast rise exponential decay (``FRED'') profiles, respectively, for the remainder of this work.

\begin{figure*}
\centering
\begin{tabular}{cc}
\includegraphics[width=0.45\linewidth,clip=true,trim=0cm 0cm 0cm 0cm]{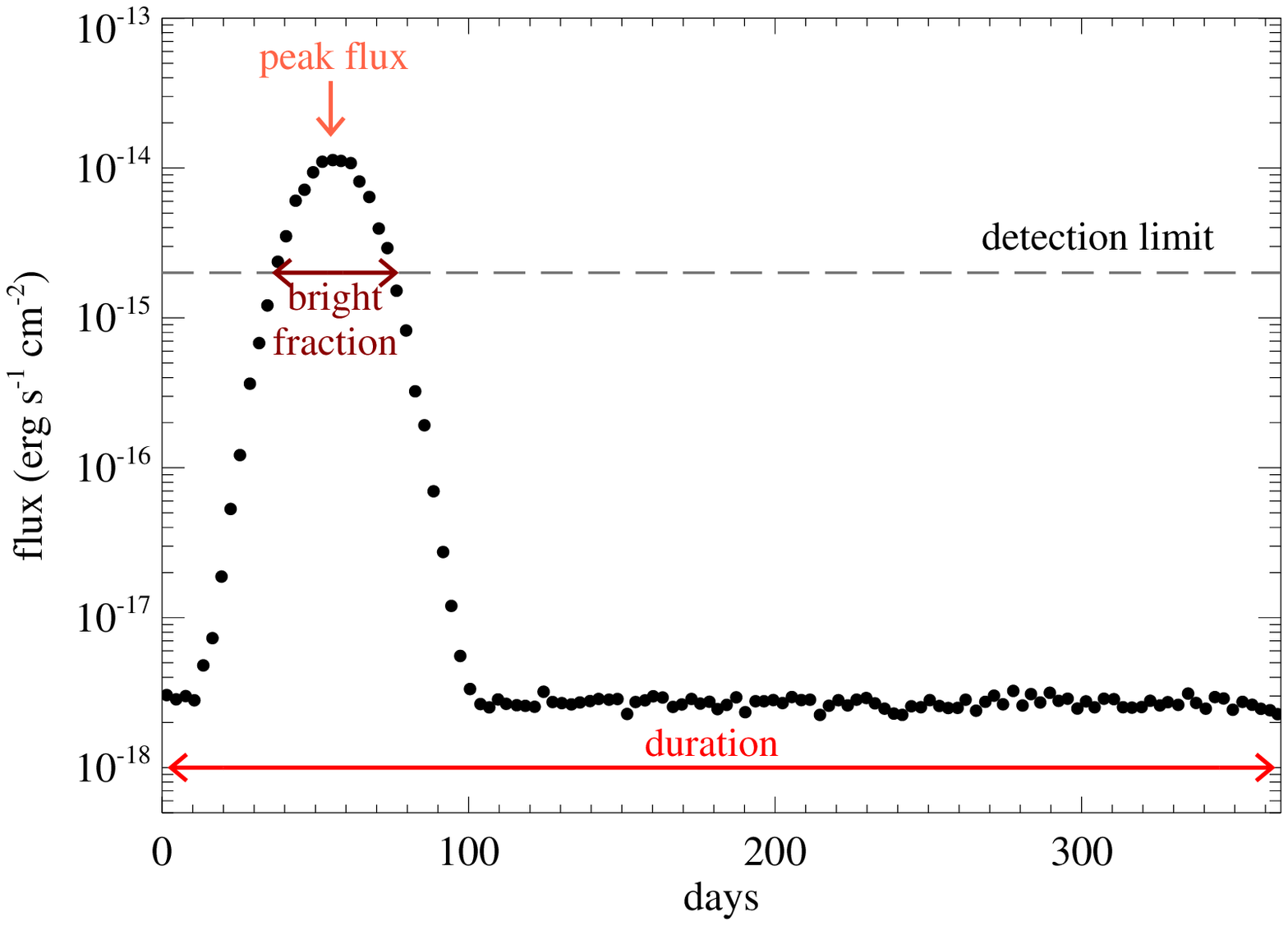} &
\includegraphics[width=0.45\linewidth,clip=true,trim=0cm 0cm 0cm 0cm]{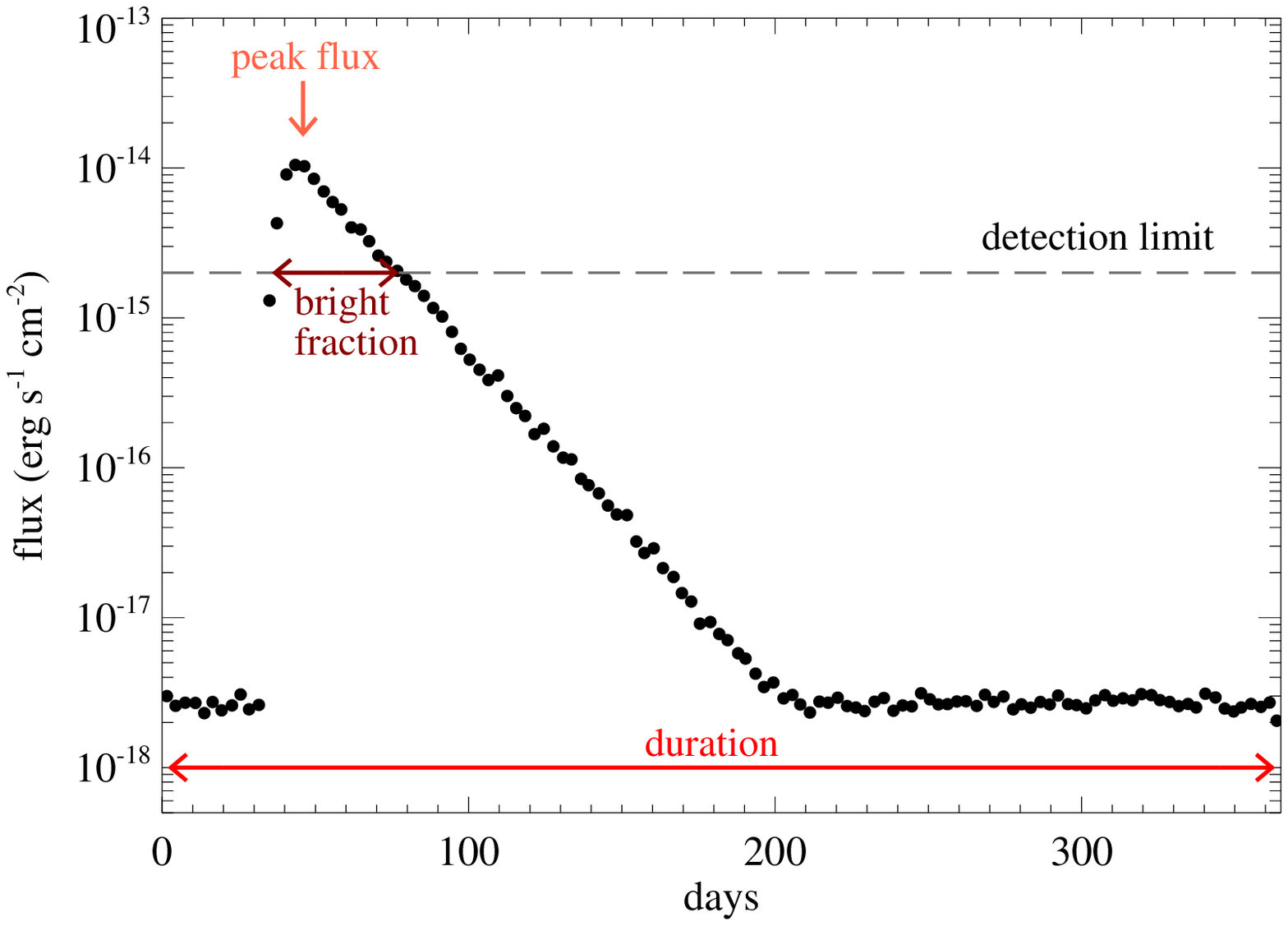} \\
\end{tabular}
\caption{Example light curve templates showing the Gaussian profile (left) and FRED profile (right). Our observational detection limit is shown by the horizontal dashed line. The duration, bright fraction, and peak flux are the model parameters that can be adjusted.} 
\label{figure_burst_profiles}
\end{figure*}

Our aim was to generate a population of synthetic XRBs (some of which followed a Gaussian profile, and some of which followed a FRED profile), and then to ``observe'' these synthetic sources and construct \lognlogs distributions in exactly the same manner as was done for our \Chandra observations. All synthetic sources were given a ``quiescent'' X-ray flux of $2\times10^{-17}$ \flux, equivalent to $\sim$10$^{33}$ \lum at the distance of NGC~300. The model light curves were described by three free parameters: the duration of time in between outbursts (hereafter referred to as the ``duration''), the fraction of time that the source spends above the 90\% limit flux of our observations (hereafter referred to as the ``bright fraction''), and the peak flux. Peak fluxes were randomly drawn from a power law distribution of fluxes above the detection limit of our survey ($\sim10^{36}$ \lum), with the highest possible flux of $2\times10^{-12}$ \flux (equivalent to 10$^{39}$ \lum). The power law index of this distribution was our free parameter. Two template burst profiles are shown in Figure~\ref{figure_burst_profiles}.

In order to determine what effect the free parameters in our profiles had on the resulting \lognlogs distributions, we generated a grid of 145 models: 125 of these used a roughly 50/50 mix of Gaussian and FRED profiles, and ten of these models were randomly selected and re-run two additional times, once using exclusively Gaussian profiles and once using exclusively FRED profiles, to generate the \lognlogs distributions. The duration parameter could have a value of 122, 183, 365, 730, or 1095 days (corresponding to a burst frequency of once per four months, six months, one year, two years, or three years), the bright fraction was set to 1\%, 10\%, 30\%, 50\%, or 70\%, and the power law index that determined the distribution from which the peak flux was drawn could have a value of 0.6 (i.e., relatively flat distribution, indicating that bright bursts are somewhat likely to occur), 1.8, 2.6, 3.0, or 3.4 (i.e., steep distributions indicating a strong preference for fainter peak fluxes). While XRBs have been observed undergo outbursts on timescales less than 122 days, the time between our observations is too long for us to place any meaningful constraints on duration parameters this short. The bright fraction is related to the X-ray source duty cycle and the number of observations; typical duty cycles of XRBs are $\sim$20--60\% \citep{Romano+14b}, but with our three observations we are not able to reliably test this currently-accepted range. For each model, 500 synthetic X-ray sources were created and three random fluxes were drawn from their light curves. Fluxes were assigned a $\sim$10-25\% uncertainty, typical to the uncertainties of faint sources in our observations. The resulting synthetic \lognlogs distributions were then fit using a broken power-law in Sherpa in an identical manner as was done for the observed variable source distributions.

\begin{figure}
\centering
\includegraphics[width=1.05\linewidth]{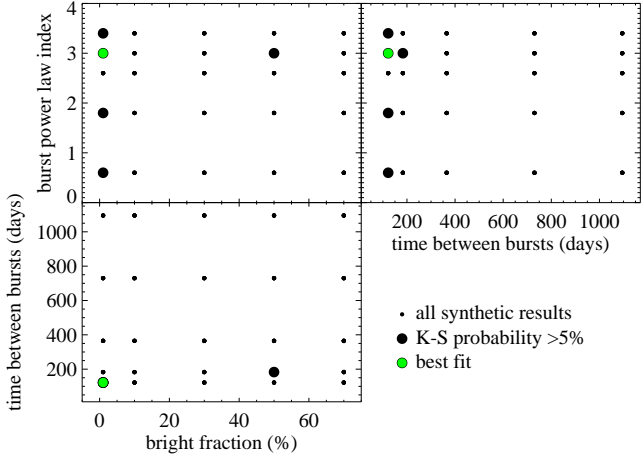} 
\caption{The parameter space explored by our synthetic \lognlogs distributions. All model runs are shown by the small dots. Larger circles indicate models that yielded K-S probabilities $>$5\%, and the green circle indicates our best-fit model.}
\label{figure_synthetic_parspace}
\end{figure}

\begin{figure}
\centering
\includegraphics[width=1\linewidth]{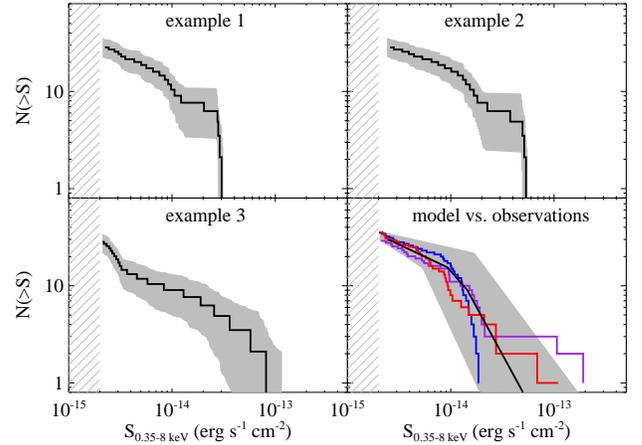} 
\caption{Examples of the \lognlogs distributions produced by our best-fit variable source model. The bottom-right panel shows the best-fit model (black line, with uncertainties shown by the shaded gray region) compared to the observed variable source distributions. ObsID 12238 is shown in blue, 16028 is shown in purple, and 16029 is shown in red. The gray-lined region indicates fluxes below our 90\% completeness limit; only sources above this limit are shown.} 
\label{figure_synthetic_examples}
\end{figure}

For each resulting fit, we record the best-fit values of $\gamma_{\rm f}$, $\gamma_{\rm b}$, and $S_{\rm b}$, and the corresponding uncertainties. We additionally performed a two-sided K-S test of each model against the observed variable source distributions. Models that returned very low K-S probabilities ($<$1\%) for all three observations are not consistent with the observations. We found no significant difference in the resulting \lognlogs distributions or fit quality when only Gaussian or FRED profiles were used compared to the 50/50 mix. Table~\ref{table_synthetic_parspace} summarizes the model parameters and resulting \lognlogs parameters for the five models that yielded K-S probabilities $>$5\%. Figure~\ref{figure_synthetic_parspace} shows the bright fraction, duration, and burst power law index parameter space we explored; large circles show models with K-S probabilities $>$5\% when compared to the observations, and the best-fit model is shown in green. The best-fit model was selected as the one with the highest K-S probability, and has a bright fraction of 1\%, a duration of 122 days (corresponding to one outburst every $\sim$4 months), and a peak flux power law index of 3.0. This model yields \lognlogs fit parameters $\gamma_{\rm f}=1.5^{+0.3}_{-0.5}$, $\gamma_{\rm b}=3.0^{+0.5}_{-0.7}$, and $S_{\rm b}=(1.2^{+0.5}_{-0.6})\times10^{-14}$ \flux. Figure~\ref{figure_synthetic_examples} provides three examples of \lognlogs distributions produced by our best-fit model, compared to the observed variable source distributions.

\begin{table*}[ht]
\centering
\caption{Best-Fit Synthetic Source Models}
\begin{tabular}{ccccccc} 
\hline \hline
Bright 		& Burst Power	& Duration	& K-S		& \multicolumn{3}{c}{\lognlogs Best-Fit Parameters}			\\ \cline{5-7}
Fraction (\%)	& Law Index	& (days)		& Probability	&$\gamma_{\rm f}$	& $\gamma_{\rm b}$	& $S_{\rm b}^a$	\\
(1)			& (2)			& (3)			& (4)			& (5)				& (6)				& (7)				\\
\hline
1		& 0.6		& 122	& 0.05		& 1.7$^{+0.5}_{-0.3}$	& 2.1$^{+0.4}_{-0.6}$	& 1.7$^{+0.7}_{-0.4}$		\\
1		& 1.8		& 122	& 0.06		& 1.6$\pm$0.3			& 5.0$^{+0.7}_{-0.5}$	& 0.8$^{+0.6}_{-0.4}$		\\
{\bf 1}	& {\bf 3.0}	& {\bf 122}	& {\bf 0.10}	& {\bf 1.5$^{+0.3}_{-0.5}$}	& {\bf 3.0$^{+0.5}_{-0.7}$}	& {\bf 1.2$^{+0.5}_{-0.6}$}		\\
1		& 3.4		& 122	& 0.09		& 1.6$^{+0.5}_{-0.3}$	& 4.8$^{+1.3}_{-0.6}$	& 0.8$^{+0.6}_{-0.4}$		\\
50		& 3.0		& 183	& 0.06		& 1.7$^{+0.5}_{-0.3}$	& 6.7$^{+0.7}_{-0.8}$	& 1.1$^{+0.3}_{-0.5}$		\\
\hline \hline
\end{tabular}
\tablecomments{Best-fit synthetic source models. The best-fit model, with the highest K-S probability compared to the observations, is shown in bold text.\newline
$^a$Break flux is given in units of 10$^{-14}$ \flux.}
\label{table_synthetic_parspace}
\end{table*}

The observed \lognlogs distributions clearly favor models with X-ray sources that burst multiple times per year. The two best models additionally have small bright fractions, 1\%, and steep burst power law indices, indicating that the majority of XRB bursts occur at relatively faint fluxes, and that bright X-ray bursts are severely underrepresented in our survey. Of the tens of thousands of synthetic X-ray sources generated, only $\sim$0.4\% produced an ``observed'' peak flux above the Eddington limit of a 1.4 \Msun neutron star (2$\times10^{38}$ \lum), and they were all the result of the flatter (burst power law index of 0.6) peak flux distributions. 

It is typically assumed that, during an outburst, an XRB produces a peak luminosity at or near its Eddington limit. We therefore generated synthetic light curves in which we {\it required} the peak flux to correspond to the Eddington limit of a 1.4 \Msun NS. The faint- and bright-end power law indices were similar to those found in our observations, but the break flux is a factor of $\sim$10 higher. The synthetic \lognlogs distributions have typical break fluxes from $\sim$5$\times10^{-14}$ \flux to $\sim10^{-13}$ \flux, equivalent to $\sim$(2--5)$\times10^{37}$ \lum at the distance of NGC~300, across {\it all} values of duration and bright fraction. The best K-S probability from the Eddington models was $\sim$5$\times10^{-7}$, indicating a significant difference from the observations. These results therefore suggest that the majority of outbursting XRBs in NGC~300 occur at sub-Eddington luminosities.

	\subsection{X-ray Binary Variability and Evolution}
The persistent sources in NGC~300 are likely dominated by background AGN, with a handful of RLOF LMXBs, while the variable source distribution is likely a mix of HMXBs and LMXBs, dominated by HMXBs, outbursting at sub-Eddington rates. Assuming a 1.4 \Msun NS, the typical variable source break flux corresponds to a luminosity of $\sim$(2--6)$\times10^{36}$ \lum corresponds to an accretion rate of 1--3\% the Eddington limit. Furthermore, we have modeled the \lognlogs distributions originating from outbursting X-ray sources, and found we can match the broken power law shape seen in our observations when the distribution of peak fluxes significantly favors faint outbursts. Although bright X-ray bursts are better studied, particularly in extragalactic sources, our results imply that they do not represent the ``typical'' X-ray variable source in NGC~300.

The faint-end of the variable source \lognlogs distribution has a power law index of $\sim$1.7, similar to the universal XLF of HMXBs. Recently, \cite{Zuo+14} suggested that the shape of the XLF could be explained by the common envelope (CE) evolution of the progenitor binary as it evolves. Their simulations were able to reproduce the observed XLF for two common formalisms of CE evolution: the $\alpha_{\rm CE}$ formalism (also called the ``energy-budget'' approach) and the $\gamma$-formalism (the ``angular momentum'' budget approach). In the $\alpha_{\rm CE}$ formalism, the CE evolution is parameterized in terms of the orbital energy ($E_{\rm orb}$) and the envelope binding energy ($E_{\rm bind}$). The parameter $\alpha_{\rm CE}$ describes the efficiency at which the system's orbital energy is converted into kinetic energy that is used to eject the envelope \citep{Webbink84,Webbink08}. Under the $\gamma$-formalism, CE evolution is instead parameterized by the ratio of the fraction of angular momentum lost during the CE phase and the fraction of mass loss in the system, and has been more successful in explaining the observed properties of double-white dwarf binaries, cataclysmic variables, and binary main sequence stars \citep{Nelemans+05}.

The simulated XRB populations in \cite{Zuo+14} included a large fraction of variable sources, including both NS and BH primaries with supergiant and Be companions undergoing either Roche lobe overflow or direct wind accretion. However, persistent and variable sources were not separated during the construction of the XLFs. The shapes of the simulated XLFs were similar for both CE evolution parameterizations, but the number of characteristics of the resulting HMXB populations were distinctly different from one another. Under the $\alpha_{\rm CE}$-formalism, XRBs with MS companions are the dominant source of X-rays below $10^{37}$ \lum, while the XRBs with He-rich companions (e.g., evolved donors that have lost a significant fraction of their outer H envelopes) are the majority constituent under the $\gamma$-formalism. Our X-ray observations indicate that the variability properties of both HMXBs and LMXBs may contribute to the shape of the XLF independently of the evolutionary history of the system. Our X-ray observations are not adequate to determine which scenario is more likely for the low-luminosity X-ray sources in NGC~300.

\section{Conclusions}\label{conclusions}
We have studied the \lognlogs distributions of X-ray sources in NGC~300 across three epochs with \Chandra down to $\sim$10$^{36}$ \lum. This is an order of magnitude fainter than a similar study of the Antennae galaxies \citep{Zezas+07}. We find that the majority of variable X-ray sources in NGC~300 have luminosities less than $\sim5\times10^{36}$ \lum, while the brightest sources exhibit persistent X-ray emission (within a factor of $\sim$2). This result may help explain why variable X-ray sources have not had a significant impact on studies of the XLF in single, ``snapshot'' exposures, as these observations frequently detect only sources above $\sim$10$^{37}$ \lum. If all significantly variable X-ray sources are faint but numerous, as implied by our observations, then their large numbers would result in similar {\it distributions} of brightness across different observations.

The persistent sources in NGC~300 are likely undergoing mass transfer via RLOF, and their flat \lognlogs distribution is consistent with that of field LMXBs \citep{Lin+15}. However, the persistent source population is likely dominated by AGN, particularly at low fluxes ($\lesssim2\times10^{-14}$ \flux). The variable source XLF is well described by a broken power law, with a faint-end power law index similar to that of HMXBs \citep{Mineo+12} with cut-off fluxes of $\sim$(0.5--1.5)$\times10^{-14}$ \flux, corresponding to a luminosity of $\sim$(2--7)$\times10^{36}$ \lum. These observations suggest that the highly variable X-ray sources in NGC~300 are wind-accreting XRBs, possibly HMXBs undergoing Type~II outbursts, although we cannot completely rule out a significant contribution from variable LMXBs to the shape of the highly variable \lognlogs distribution.

We were able to reproduce the observed \lognlogs distributions of variable sources in NGC~300 by assuming generic profiles of the X-ray outbursts, with no assumptions made about the prior evolutionary history of the systems. It is unclear to what degree the shape of the XLF is the result of CE evolution, and how much may be driven by the variability properties of the underlying source distribution. CE evolution likely plays a key role in determining the shape of the persistent-source XLF and determining the fraction of the massive binary population that later produces variable XRBs. A better understanding of how the X-ray outburst profiles are related to its prior evolutionary history may provide clues to the missing link between the variability properties of XRBs and the ``universal'' shape of the HMXB XLF; repeated \Chandra observations of nearby, star-forming galaxies down to $\sim10^{36}$ \lum are necessary to better constrain the XLF shape of outbursting and RLOF XRBs.

\acknowledgements
We would like to thank the anonymous referee for the constructive feedback that improved this manuscript. Support for this work was provided by the National Aeronautics and Space Administration through \Chandra Award Number G04-15088X issued by the \Chandra X-ray Observatory Center, which is operated by the Smithsonian Astrophysical Observatory for and on behalf of the National Aeronautics Space Administration under contract NAS8-03060. PPP and TJG acknowledge support under NASA contract NAS8-03060. This research has made use of software provided by the \Chandra X-ray Center (CXC) in the application packages CIAO, ChIPS, and Sherpa, and the NASA/IPAC Extragalactic Database (NED) which is operated by the Jet Propulsion Laboratory, California Institute of Technology, under contract with the National Aeronautics and Space Administration.

\bibliography{apjmnemonic,ms}

\begin{thebibliography}{}

\bibitem[\protect\citeauthoryear{{Antoniou} et~al.}{{Antoniou}
  et~al.}{2010}]{Antoniou+10}
{Antoniou}, V., {Zezas}, A., {Hatzidimitriou}, D.,  \& {Kalogera}, V. 2010,
  \apjl, 716, L140

\bibitem[\protect\citeauthoryear{{Asai} et~al.}{{Asai} et~al.}{2012}]{Asai+12}
{Asai}, K., et~al. 2012, \pasj, 64

\bibitem[\protect\citeauthoryear{{Binder} et~al.}{{Binder}
  et~al.}{2015}]{Binder+15}
{Binder}, B., {Gross}, J., {Williams}, B.~F.,  \& {Simons}, D. 2015, \mnras,
  451, 4471

\bibitem[\protect\citeauthoryear{{Binder} et~al.}{{Binder}
  et~al.}{2012}]{Binder+12}
{Binder}, B., et~al. 2012, \apj, 758, 15

\bibitem[\protect\citeauthoryear{{Binder} et~al.}{{Binder}
  et~al.}{2011}]{Binder+11}
{Binder}, B., {Williams}, B.~F., {Eracleous}, M., {Garcia}, M.~R., {Anderson},
  S.~F.,  \& {Gaetz}, T.~J. 2011, \apj, 742, 128

\bibitem[\protect\citeauthoryear{{Bland-Hawthorn} et~al.}{{Bland-Hawthorn}
  et~al.}{2005}]{Bland-Hawthorn+05}
{Bland-Hawthorn}, J., {Vlaji{\'c}}, M., {Freeman}, K.~C.,  \& {Draine}, B.~T.
  2005, \apj, 629, 239

\bibitem[\protect\citeauthoryear{{Bondi} \& {Hoyle}}{{Bondi} \&
  {Hoyle}}{1944}]{Bondi+44}
{Bondi}, H.,  \& {Hoyle}, F. 1944, \mnras, 104, 273

\bibitem[\protect\citeauthoryear{{Cappelluti} et~al.}{{Cappelluti}
  et~al.}{2009}]{Cappelluti+09}
{Cappelluti}, N., et~al. 2009, \aap, 497, 635

\bibitem[\protect\citeauthoryear{{Dalcanton} et~al.}{{Dalcanton}
  et~al.}{2009}]{Dalcanton+09}
{Dalcanton}, J.~J., et~al. 2009, \apjs, 183, 67

\bibitem[\protect\citeauthoryear{{Davidson} \& {Ostriker}}{{Davidson} \&
  {Ostriker}}{1973}]{Davidson+73}
{Davidson}, K.,  \& {Ostriker}, J.~P. 1973, \apj, 179, 585

\bibitem[\protect\citeauthoryear{{Doe} et~al.}{{Doe} et~al.}{2007}]{Doe+07}
{Doe}, S., et~al. 2007, in Astronomical Society of the Pacific Conference
  Series, Vol. 376, Astronomical Data Analysis Software and Systems XVI, ed.
  R.~A. {Shaw}, F.~{Hill}, \& D.~J. {Bell}, 543

\bibitem[\protect\citeauthoryear{{Ducci} et~al.}{{Ducci}
  et~al.}{2014}]{Ducci+14}
{Ducci}, L., {Doroshenko}, V., {Romano}, P., {Santangelo}, A.,  \& {Sasaki}, M.
  2014, \aap, 568, A76

\bibitem[\protect\citeauthoryear{{Freeman}, {Doe}, \&
  {Siemiginowska}}{{Freeman} et~al.}{2001}]{Freeman+01}
{Freeman}, P., {Doe}, S.,  \& {Siemiginowska}, A. 2001, in \procspie, Vol.
  4477, Astronomical Data Analysis, ed. J.-L. {Starck} \& F.~D. {Murtagh}, 76

\bibitem[\protect\citeauthoryear{{Freeman} et~al.}{{Freeman}
  et~al.}{2002}]{Freeman+02}
{Freeman}, P.~E., {Kashyap}, V., {Rosner}, R.,  \& {Lamb}, D.~Q. 2002, \apjs,
  138, 185

\bibitem[\protect\citeauthoryear{{Gehrels}}{{Gehrels}}{1986}]{Gehrels86}
{Gehrels}, N. 1986, \apj, 303, 336

\bibitem[\protect\citeauthoryear{{Georgakakis} et~al.}{{Georgakakis}
  et~al.}{2008}]{Georgakakis+08}
{Georgakakis}, A., {Nandra}, K., {Laird}, E.~S., {Aird}, J.,  \& {Trichas}, M.
  2008, \mnras, 388, 1205

\bibitem[\protect\citeauthoryear{{Gogarten} et~al.}{{Gogarten}
  et~al.}{2010}]{Gogarten+10}
{Gogarten}, S.~M., et~al. 2010, \apj, 712, 858

\bibitem[\protect\citeauthoryear{{Grimm}, {Gilfanov}, \& {Sunyaev}}{{Grimm}
  et~al.}{2003}]{Grimm+03}
{Grimm}, H.-J., {Gilfanov}, M.,  \& {Sunyaev}, R. 2003, \mnras, 339, 793

\bibitem[\protect\citeauthoryear{{Jeltema} et~al.}{{Jeltema}
  et~al.}{2003}]{Jeltema+03}
{Jeltema}, T.~E., {Canizares}, C.~R., {Buote}, D.~A.,  \& {Garmire}, G.~P.
  2003, \apj, 585, 756

\bibitem[\protect\citeauthoryear{{Kalberla} et~al.}{{Kalberla}
  et~al.}{2005}]{Kalberla+05}
{Kalberla}, P.~M.~W., {Burton}, W.~B., {Hartmann}, D., {Arnal}, E.~M.,
  {Bajaja}, E., {Morras}, R.,  \& {P{\"o}ppel}, W.~G.~L. 2005, \aap, 440, 775

\bibitem[\protect\citeauthoryear{{Kilgard} et~al.}{{Kilgard}
  et~al.}{2002}]{Kilgard+02}
{Kilgard}, R.~E., {Kaaret}, P., {Krauss}, M.~I., {Prestwich}, A.~H., {Raley},
  M.~T.,  \& {Zezas}, A. 2002, \apj, 573, 138

\bibitem[\protect\citeauthoryear{{Kim} et~al.}{{Kim} et~al.}{2004}]{Kim+04x}
{Kim}, D.-W., et~al. 2004, \apjs, 150, 19

\bibitem[\protect\citeauthoryear{{Kim} et~al.}{{Kim} et~al.}{2009}]{Kim+09}
{Kim}, D.-W., et~al. 2009, \apj, 703, 829

\bibitem[\protect\citeauthoryear{{Lamers}, {van den Heuvel}, \&
  {Petterson}}{{Lamers} et~al.}{1976}]{Lamers+76}
{Lamers}, H.~J.~G.~L.~M., {van den Heuvel}, E.~P.~J.,  \& {Petterson}, J.~A.
  1976, \aap, 49, 327

\bibitem[\protect\citeauthoryear{{Larsen} \& {Richtler}}{{Larsen} \&
  {Richtler}}{1999}]{Larsen+99}
{Larsen}, S.~S.,  \& {Richtler}, T. 1999, \aap, 345, 59

\bibitem[\protect\citeauthoryear{{Laycock} et~al.}{{Laycock}
  et~al.}{2005}]{Laycock+05}
{Laycock}, S., {Corbet}, R.~H.~D., {Coe}, M.~J., {Marshall}, F.~E.,
  {Markwardt}, C.,  \& {Lochner}, J. 2005, \apjs, 161, 96

\bibitem[\protect\citeauthoryear{{Laycock} et~al.}{{Laycock}
  et~al.}{2010}]{Laycock+10}
{Laycock}, S., {Zezas}, A., {Hong}, J., {Drake}, J.~J.,  \& {Antoniou}, V.
  2010, \apj, 716, 1217

\bibitem[\protect\citeauthoryear{{Lehmer} et~al.}{{Lehmer}
  et~al.}{2010}]{Lehmer+10}
{Lehmer}, B.~D., {Alexander}, D.~M., {Bauer}, F.~E., {Brandt}, W.~N.,
  {Goulding}, A.~D., {Jenkins}, L.~P., {Ptak}, A.,  \& {Roberts}, T.~P. 2010,
  \apj, 724, 559

\bibitem[\protect\citeauthoryear{{Lehmer} et~al.}{{Lehmer}
  et~al.}{2014}]{Lehmer+14}
{Lehmer}, B.~D., et~al. 2014, \apj, 789, 52

\bibitem[\protect\citeauthoryear{{Lewin}, {van Paradijs}, \& {van den
  Heuvel}}{{Lewin} et~al.}{1997}]{Lewin+97}
{Lewin}, W.~H.~G., {van Paradijs}, J.,  \& {van den Heuvel}, E.~P.~J. 1997,
  {X-ray Binaries} 674

\bibitem[\protect\citeauthoryear{{Lin} et~al.}{{Lin} et~al.}{2015}]{Lin+15}
{Lin}, D., et~al. 2015, \apj, 808, 20

\bibitem[\protect\citeauthoryear{{Liu}}{{Liu}}{2011}]{Liu11}
{Liu}, J. 2011, \apjs, 192, 10

\bibitem[\protect\citeauthoryear{{Liu}, {van Paradijs}, \& {van den
  Heuvel}}{{Liu} et~al.}{2006}]{Liu+06}
{Liu}, Q.~Z., {van Paradijs}, J.,  \& {van den Heuvel}, E.~P.~J. 2006, \aap,
  455, 1165

\bibitem[\protect\citeauthoryear{{Lutovinov} et~al.}{{Lutovinov}
  et~al.}{2013}]{Lutovinov+13}
{Lutovinov}, A.~A., {Revnivtsev}, M.~G., {Tsygankov}, S.~S.,  \& {Krivonos},
  R.~A. 2013, \mnras, 431, 327

\bibitem[\protect\citeauthoryear{{Maccarone}}{{Maccarone}}{2003}]{Maccarone03}
{Maccarone}, T.~J. 2003, \aap, 409, 697

\bibitem[\protect\citeauthoryear{{McSwain} \& {Gies}}{{McSwain} \&
  {Gies}}{2005}]{McSwain+05}
{McSwain}, M.~V.,  \& {Gies}, D.~R. 2005, \apjs, 161, 118

\bibitem[\protect\citeauthoryear{{Mineo}, {Gilfanov}, \& {Sunyaev}}{{Mineo}
  et~al.}{2012}]{Mineo+12}
{Mineo}, S., {Gilfanov}, M.,  \& {Sunyaev}, R. 2012, \mnras, 419, 2095

\bibitem[\protect\citeauthoryear{{Mu{\~n}oz-Mateos} et~al.}{{Mu{\~n}oz-Mateos}
  et~al.}{2007}]{Munoz+07}
{Mu{\~n}oz-Mateos}, J.~C., {Gil de Paz}, A., {Boissier}, S., {Zamorano}, J.,
  {Jarrett}, T., {Gallego}, J.,  \& {Madore}, B.~F. 2007, \apj, 658, 1006

\bibitem[\protect\citeauthoryear{{Negueruela} et~al.}{{Negueruela}
  et~al.}{2008}]{Negueruela+08}
{Negueruela}, I., {Torrej{\'o}n}, J.~M., {Reig}, P., {Rib{\'o}}, M.,  \&
  {Smith}, D.~M. 2008, in American Institute of Physics Conference Series, Vol.
  1010, A Population Explosion: The Nature \& Evolution of X-ray Binaries in
  Diverse Environments, ed. R.~M. {Bandyopadhyay}, S.~{Wachter}, D.~{Gelino},
  \& C.~R. {Gelino}, 252

\bibitem[\protect\citeauthoryear{{Nelemans} \& {Tout}}{{Nelemans} \&
  {Tout}}{2005}]{Nelemans+05}
{Nelemans}, G.,  \& {Tout}, C.~A. 2005, \mnras, 356, 753

\bibitem[\protect\citeauthoryear{{Paolillo} et~al.}{{Paolillo}
  et~al.}{2004}]{Paolillo+04}
{Paolillo}, M., {Schreier}, E.~J., {Giacconi}, R., {Koekemoer}, A.~M.,  \&
  {Grogin}, N.~A. 2004, \apj, 611, 93

\bibitem[\protect\citeauthoryear{{Peacock} \& {Zepf}}{{Peacock} \&
  {Zepf}}{2016}]{Peacock+16}
{Peacock}, M.~B.,  \& {Zepf}, S.~E. 2016, \apj, 818, 33

\bibitem[\protect\citeauthoryear{{Reig}}{{Reig}}{2008}]{Reig08}
{Reig}, P. 2008, \aap, 489, 725

\bibitem[\protect\citeauthoryear{{Reig}}{{Reig}}{2011}]{Reig11}
{Reig}, P. 2011, \apss, 332, 1

\bibitem[\protect\citeauthoryear{{Reig} \& {Nespoli}}{{Reig} \&
  {Nespoli}}{2013}]{Reig+13}
{Reig}, P.,  \& {Nespoli}, E. 2013, \aap, 551, A1

\bibitem[\protect\citeauthoryear{{Romano} et~al.}{{Romano}
  et~al.}{2014a}]{Romano+14}
{Romano}, P., {Ducci}, L., {Mangano}, V., {Esposito}, P., {Bozzo}, E.,  \&
  {Vercellone}, S. 2014a, \aap, 568, A55

\bibitem[\protect\citeauthoryear{{Romano} et~al.}{{Romano}
  et~al.}{2014b}]{Romano+14b}
{Romano}, P., {Guidorzi}, C., {Segreto}, A., {Ducci}, L.,  \& {Vercellone}, S.
  2014b, \aap, 572, A97

\bibitem[\protect\citeauthoryear{{Romano} et~al.}{{Romano}
  et~al.}{2011}]{Romano+11}
{Romano}, P., et~al. 2011, \mnras, 410, 1825

\bibitem[\protect\citeauthoryear{{Shtykovskiy} \& {Gilfanov}}{{Shtykovskiy} \&
  {Gilfanov}}{2005}]{Shty+Gilfanov05}
{Shtykovskiy}, P.,  \& {Gilfanov}, M. 2005, \mnras, 362, 879

\bibitem[\protect\citeauthoryear{{Shtykovskiy} \& {Gilfanov}}{{Shtykovskiy} \&
  {Gilfanov}}{2007}]{Shty+07}
{Shtykovskiy}, P.~E.,  \& {Gilfanov}, M.~R. 2007, Astronomy Letters, 33, 437

\bibitem[\protect\citeauthoryear{{Sidoli} et~al.}{{Sidoli}
  et~al.}{2008}]{Sidoli+08}
{Sidoli}, L., et~al. 2008, \apj, 687, 1230

\bibitem[\protect\citeauthoryear{{Soldi} et~al.}{{Soldi}
  et~al.}{2014}]{Soldi+14}
{Soldi}, S., et~al. 2014, \aap, 563, A57

\bibitem[\protect\citeauthoryear{{Stroh} \& {Falcone}}{{Stroh} \&
  {Falcone}}{2013}]{Stroh+13}
{Stroh}, M.~C.,  \& {Falcone}, A.~D. 2013, \apjs, 207, 28

\bibitem[\protect\citeauthoryear{{Walter} et~al.}{{Walter}
  et~al.}{2015}]{Walter+15}
{Walter}, R., {Lutovinov}, A.~A., {Bozzo}, E.,  \& {Tsygankov}, S.~S. 2015,
  \aapr, 23, 2

\bibitem[\protect\citeauthoryear{{Webbink}}{{Webbink}}{1984}]{Webbink84}
{Webbink}, R.~F. 1984, \apj, 277, 355

\bibitem[\protect\citeauthoryear{{Webbink}}{{Webbink}}{2008}]{Webbink08}
{Webbink}, R.~F. 2008, in Astrophysics and Space Science Library, Vol. 352,
  Astrophysics and Space Science Library, ed. E.~F. {Milone}, D.~A. {Leahy}, \&
  D.~W. {Hobill}, 233

\bibitem[\protect\citeauthoryear{{Williams} et~al.}{{Williams}
  et~al.}{2013}]{Williams+13}
{Williams}, B.~F., {Binder}, B.~A., {Dalcanton}, J.~J., {Eracleous}, M.,  \&
  {Dolphin}, A. 2013, \apj, 772, 12

\bibitem[\protect\citeauthoryear{{Zezas} et~al.}{{Zezas}
  et~al.}{2007}]{Zezas+07}
{Zezas}, A., {Fabbiano}, G., {Baldi}, A., {Schweizer}, F., {King}, A.~R.,
  {Rots}, A.~H.,  \& {Ponman}, T.~J. 2007, \apj, 661, 135

\bibitem[\protect\citeauthoryear{{Zuo} \& {Li}}{{Zuo} \& {Li}}{2014}]{Zuo+14}
{Zuo}, Z.-Y.,  \& {Li}, X.-D. 2014, \apj, 797, 45

\end{thebibliography}
\bibliographystyle{apj}
                                                                                             
\end{document}